\documentclass[structabstract]{aa}
\usepackage{graphicx,amssymb,natbib,amsmath,txfonts,cite}
\bibpunct{(}{)}{;}{a}{}{,}
\def\msun{{\rm M}_\odot}

\begin{document}

\title{The population of Young Stellar Clusters throughout the disk of M33}

    \author{Saurabh Sharma\inst{1,2}, Edvige Corbelli\inst{1}, 
Carlo Giovanardi\inst{1}, Leslie K. Hunt\inst{1}
          \and
          Francesco Palla\inst{1} 
          }

   \institute{INAF-Osservatorio Astrofisico di Arcetri,
	Largo E. Fermi 5, 50125 Firenze, Italy\\
              \email{saurabh;edvige;giova;hunt;palla@arcetri.astro.it}
         \and
             Aryabhatta Research Institute of Observational Sciences (ARIES), 
Manora Peak, Nainital, 263 129, India\\
             \email{saurabh@aries.res.in} 
             }

\date{Received ....... .., 2011; accepted ..... .., 2011}

\abstract
{}
{The properties of young stellar clusters (YSCs) in M33, identified
from the center out to about twice the size of the bright star-forming disk,
are investigated to determine possible spatial 
and time variations of the star formation process 
in this Local Group blue galaxy.}
{915 MIR sources have been extracted from the Spitzer $24\,\mu$m image. 
Upon inspection of H$\alpha$ and GALEX 
images and exclusion of evolved AGB stars, a sample of 648 objects is selected 
as candidate YSCs and their luminosity function is examined.  
The spectral energy distribution of each object, based on aperture 
photometry, is compared with Starburst99 models to derive age, 
mass and $A_V$ of individual clusters. 
In the analysis we allow for different values of the upper mass cutoff of the 
stellar initial mass function (IMF), the porosity of the ISM, 
and the dustiness of HII regions. 
We also examine the influence of different dust models and include corrections for 
incompleteness of the IMF.} 
{We find discrete MIR sources as far as the extent of the warped HI disk, i.e. 16\,kpc 
from the galaxy center. Their surface 
density has a steep radial decline beyond 4.5\,kpc, 
and flattens out beyond the optical radius at 8.5\,kpc. We are able to identify YSCs 
out to 12~kpc. At large galactocentric radii, 
the paucity of luminous clusters and the relevance of hot dust emission 
become evident from the analysis of the bolometric and MIR luminosity functions. 
The YSC mass and size are correlated with a log-log slope of 
$2.09\pm0.01$, similar to that measured for giant molecular clouds 
in M33 and the Milky Way, which represent the protocluster environment.
Most of the YSCs in our sample have $A_V\sim$0-1~mag 
and ages between 3 and 10 Myr. 
In the inner regions of M33 the clusters span a wide range of mass
$(10^2<~M~<~3\times~10^5\,\msun)$ and luminosity 
$(10^{38}<~L_{bol}<~3\times~10^{41}$~erg~s$^{-1})$, while at galactocentric radii
larger than $\sim 4$\,kpc we find a deficiency of massive clusters. 
Beyond 7~kpc, where the H$\alpha$ surface brightness drops significantly, 
the dominant YSC population has $M<10^3\,\msun$ and a slightly older age (10~Myrs). 
This implies the occurrence of star formation events about 10\,Myr ago as far as 10-12\,kpc 
from the center of M33.
The cluster $L_{FUV}~vs.~L_{H\alpha}$ relation is non-linear 
for $L_{FUV}<10^{39}$\,erg\,s$^{-1}$, in agreement with randomly 
sampled models of the IMF which, furthermore, 
shows no appreciable variation throughout the M33 disk.} 
{}
 
\authorrunning{Saurabh et al.}
\titlerunning{Young clusters in M33}
\maketitle

\section{Introduction}

Young stellar clusters (YSCs) are important tools for the study of the star
formation (SF) process in galaxies. Having just formed from
gravitationally bound molecular clouds and often still embedded in the parent nebular regions,
they are a unique laboratory for understanding the early dynamical evolution of the associated 
stellar population, the properties of the Initial Mass Function (IMF), 
and the interaction between young massive stars and the interstellar medium.
Massive YSCs and complexes contain hot massive stars which play an 
important role in galaxy evolution,
by ionizing, heating and enriching the ISM of heavy elements. 
They power the hot infrared (IR) radiation
of dust and, at the end of their lives, they eventually trigger further SF events 
\citep{2010A&A...521A..41G}.
The IMF is a probabilistic function describing the distribution in 
mass of stars at their birth; 
its logarithmic slope above $\sim$1~$\msun$ has been often found close to the original 
value of -2.3 \citep{1955ApJ...121..161S}.
Given the rarity of massive stars predicted by the IMF, a statistically 
meaningful ensemble of YSCs is needed to test the IMF at its upper end. 

External galaxies offer large numbers and a variety of SF sites but,
except for the Milky Way satellite galaxies, YSCs are too compact for individual
stars to be resolved. 
Even though the physical scales which can be spatially resolved will never be as
small as in Galactic surveys, some global properties of the YSCs can be investigated
using their integrated spectral energy distribution, from the UV to  long enough
wavelengths to overcome extinction \citep{2010A&A...521A..41G}.  
Tracers of SF such as IR, H$\alpha$, far-ultraviolet (FUV), and near-ultraviolet 
(NUV) emission make a good combination to study the properties of YSCs.
For example, \citet{2009A&A...495..479C} have probed the stochasticity of the IMF using 
multiwavelength photometry of a few tens of infrared selected YSCs in the inner disk of M33.
In addition to the much reduced distance ambiguities,
the advantage of studying YSCs in external galaxies is that 
large samples can be easily selected to probe different environments across the disk,
such as spiral arms and inter-arms, central areas versus outskirts, 
and to compare star-forming sites
in galaxies of different morphological types in a quiescent or dynamical stage.

The Local Group galaxy M33 is a quiescent blue galaxy at a  moderate inclination
showing no sign of recent mergers or interaction. 
Located at a distance D=840~kpc, it is currently forming stars
at a rate of $0.45\,\msun$\,yr$^{-1}$. 
The star formation rate (SFR) inferred from the FUV emission is slightly
higher ($0.55\,\msun$\,yr$^{-1}$) than that deduced by 
H$\alpha$ emission ($0.35\,\msun$\,yr$^{-1}$), but
the difference could be simply due to the uncertain correction for extinction.
When traced by the H$\alpha$ emission \citep{1989ApJ...344..685K,2009A&A...493..453V}, 
the SFR is found to decline radially with a scale length of about 2~kpc and then it drops sharply 
at galactocentric distances of about 7~kpc. 
This location is often referred to as the star formation edge of the disk. 
This sharp drop is less evident at other wavelengths, such as the mid-IR. 
In the far-UV the contribution of discrete sources drops at the SF edge but the diffuse emission 
has no apparent truncation or
steeper decline out to the extent of the GALEX maps \citep{2005ApJ...619L..67T}. 
The difference in the behavior traced by these indicators
can be explained in terms of a SFR which has substantially decreased during the last 
100-200~Myr, or by the lack of massive stars at large galactocentric radii. 
The former possibility has received some support by the recent deep 
optical surveys \citep{2011arXiv1106.4704G, 2011arXiv1107.0077D}
of regions beyond the optical radius where the average gas column density is 
only $2\times 10^{20}$~cm$^{-2}$ \citep{2003MNRAS.342..199C}. 
These studies have revealed the presence of a population of relatively young main sequence
stars with ages of about 100-200~Myr, 
indicating the existence of a more extended star forming disk in the recent past. 
Concentrations of ultraviolet light seen at large radii
in some nearby spiral and dwarf galaxies are often sign of in situ star-formation 
\citep{2011arXiv1107.5587H,2008ASPC..396..197G,2007ApJS..173..538T}. Localized density enhancements  
beyong the bright star-forming disk may occur and give birth to a population of low-mass clusters.
Clustering on sub-kiloparsec scale seen among stars with ages 10~Myrs close to the optical edge of 
M33 has given evidence of recent star-formation \citep{2011ApJ...728L..23D}.

Inside the SF edge, the distribution of the ISM in the disk of M33 is not at all uniform.  
The flocculent spiral arms fade away
around 4 kpc and this dramatic change might be responsible for the variation in the properties of the interstellar 
clouds. Full imaging of the molecular
cloud complexes in M33 \citep{2003ApJS..149..343E, 2004ApJ...602..723H, 2010A&A...522A...3G} has shown
that those with masses above $10^5\,\msun$ are very rare beyond 4~kpc.
\citet{2007MNRAS.379.1302B} also find that the number of young stellar groups 
also drops suddenly at radii $>$ 4~kpc. However, a population of small mass molecular clouds exists at larger 
distances and it becomes dominant. This is also shown by the positive results of deep CO searches around dim IR 
sources close to the SF edge \citep[see also][]{2011A&A...528A.116C}.

Mid-IR maps are generally very helpful to select stellar
clusters in their early evolutionary phase due to the presence of hot dust in the surrounding parent gas.
The resolution (6~arcmin) and extent (16~kpc) of the 24~$\mu$m Spitzer map of \citet{2007A&A...476.1161V}
is ideal for selecting YSC candidates throughout the whole disk of M33.
The presence of UV or H$\alpha$ emission is quite valuable to confirm 
the nature of mid-IR sources as YSC candidates since infrared emission alone cannot distinguish between YSCs and
evolved stars, such as AGBs, that also contribute to the source
counts in a galaxy \citep{2011A&A...528A.116C}. The spatial extent of UV and H$\alpha$ maps will confine our
YSC analysis to somewhat smaller radii. 

Previous studies of infrared selected YSCs in M33 have been restricted to a few tens of them located in
the inner disk. A more complete study of the properties of the whole YSC population throughout the star-forming
disk and possibly beyond it, still need to be done.   Finding YSCs   
in the outermost regions of M33 will definitively prove the in situ origin of the pervasive stellar 
population of ages 100~Myrs recently discovered.  
The main aim of this paper is to investigate the properties of YSC population in M33 
identified on the basis of their mid-IR, UV
and H$\alpha$ emission from the galaxy center out to about 12~kpc. 
By comparing the emission of YSCs in the mid-IR, H$\alpha$,
FUV and NUV bands with that predicted by theoretical models, 
we investigate possible differences in the distribution of mass, age,
IMF and extinction throughout the M33 disk. 
This will help set further constraints on the spatial and time variation of the
star formation process in this nearby blue galaxy.

The paper is organized as follows. 
In Section 2 we present the sample and the multiwavelength dataset. In Section
3 we analyze the properties of the infrared and bolometric luminosity functions.
The fitting method to the spectral energy distribution (SED) of YSCs  
is described in Section 4. 
In Section 5 we discuss the implications of the YSCs analysis on the IMF 
throughout the M33 disk. The cluster properties are summarized in Section 6  
and Section 7 lists our main conclusions.

\section{Multiwavelength data and the $24\,\mu$m catalog}

We select the young stellar cluster candidates of M33 using the $24\,\mu$m
images from Spitzer satellite data assembled by \citet{2007A&A...476.1161V}.
These authors cataloged
515 mid-infrared sources spanning over 4 orders of magnitude in luminosity
and reaching, on the faint side, values equivalent to a single B2V star.
However, this catalog is confined to the inner disk M33 with only few sources extending as
far as 7~kpc from the center. In order to extend the census of YSCs in the outermost regions of M33,
we have used the largest available $24\,\mu$m map
and the SExtractor software \citep{1996A&AS..117..393B}
to identify discrete sources following the same method of \citet{2007A&A...476.1161V} .

\begin{figure*}
\centering
\includegraphics[]{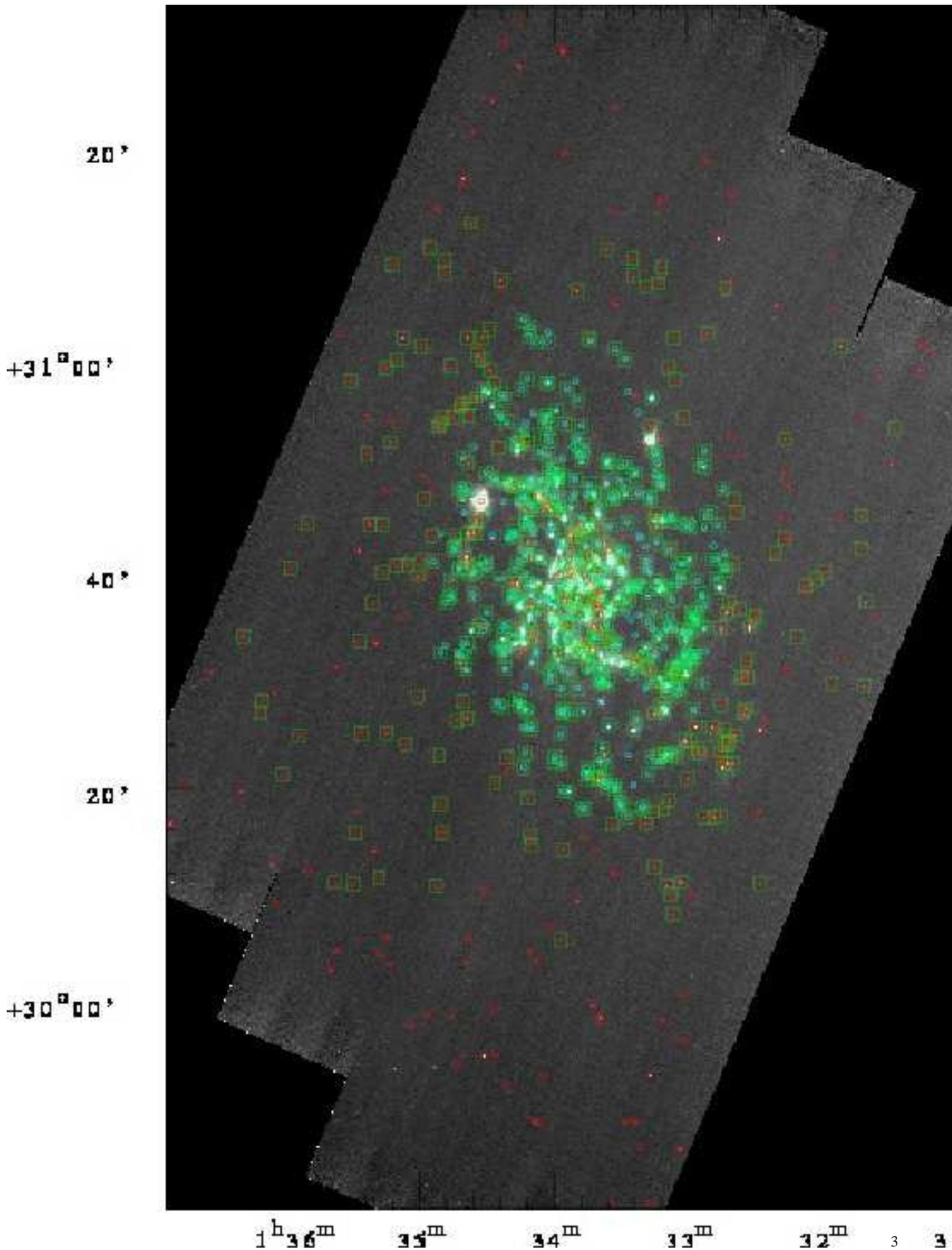}
\caption{\label{spa}  The spatial distribution of 915 Mid-IR sources (circles)
and 648 young stellar clusters (squares) overlaid on the $24\,\mu$m Spitzer image. 
The light-blue and red circles in the on-line version represent IR sources from the  \citet{2007A&A...476.1161V} catalogue
and from the present study, respectively.
}
\end{figure*}

The sources were extracted above the 8$\sigma$ level and
their photometry computed using the parameter
"flux iso" in SExtractor, which uses isophotal photometry
(sum of all pixels above threshold within the 8$\sigma$ isophote);
the background for the SExtractor photometry was the local median
in a 32x32 pixel box centered on the source.
We found 400 additional sources, so that our final list of $24\,\mu$m
selected sources amounts to 915 objects ($515+400$), whose spatial
distribution is shown in
Fig. \ref{spa} overlaid on the $24\,\mu$m mosaic.
It is evident that most of the \citet{2007A&A...476.1161V}  sources (blue circles)
lie in the inner region,
whereas those identified by us (red circles)
occupy mainly the outer regions, extending as far as 50~arcmin
from the center.
Table \ref{T515} gives the position and size of
the sources in  \citet{2007A&A...476.1161V} identified as YSCs. To the information already
present in \citet{2007A&A...476.1161V} we add the source size and its galactocentric radius. 
Table \ref{T400} refers to the newly added 400 sources;
the reported size is the semi-major axis of the ellipse fitted to the
8$\sigma$ isophote.

In order to identify YSC candidates from the selected mid-IR sources, we
complement the Spitzer $24\,\mu$m data with observations in other bandpasses:

i) Mid (8~$\mu$m) and far-infrared (70 and 160~$\mu$m) data provided by the Spitzer InfraRed Camera (IRAC) and
by the Multiband Imaging Photometer for Spitzer, as described in  \citet{2007A&A...476.1161V}.

ii) FUV and NUV maps obtained from the Galaxy Evolution Explorer (GALEX) satellite (\citet{2007ApJS..173..185G},
using the GALEX-UV images at their original sampling of $1.5$~arcsec/pixel.

iii) The narrow-band H$\alpha$ map of \citet{1998ApJ...506..135G}, described in detail
by \citet{2000ApJ...541..597H}.

We perform photometry in these other bandpasses using the position and size of the sources included
in the $24\,\mu$m catalogue with the IRAF aperture photometry {\it DAOPHOT} package.
Since the position of the source center and the extension of the emission may change
with wavelength,  
we have chosen a slightly larger photometric aperture, namely $1.5$ times the $24\,\mu$m size
as determined by SExtractor. We
estimated the background (using the ``mode'' function) in a 5-pixel wide circular annulus.
Pixels 7$\sigma$ below the background were rejected and the measurements
were corrected for foreground Galactic extinction following \citet{1998ApJ...500..525S}.

\subsection{The sample of Young Stellar Clusters}

\begin{figure*}
\centering
\hbox{
\includegraphics[height=7cm,width=7cm,angle=-90]{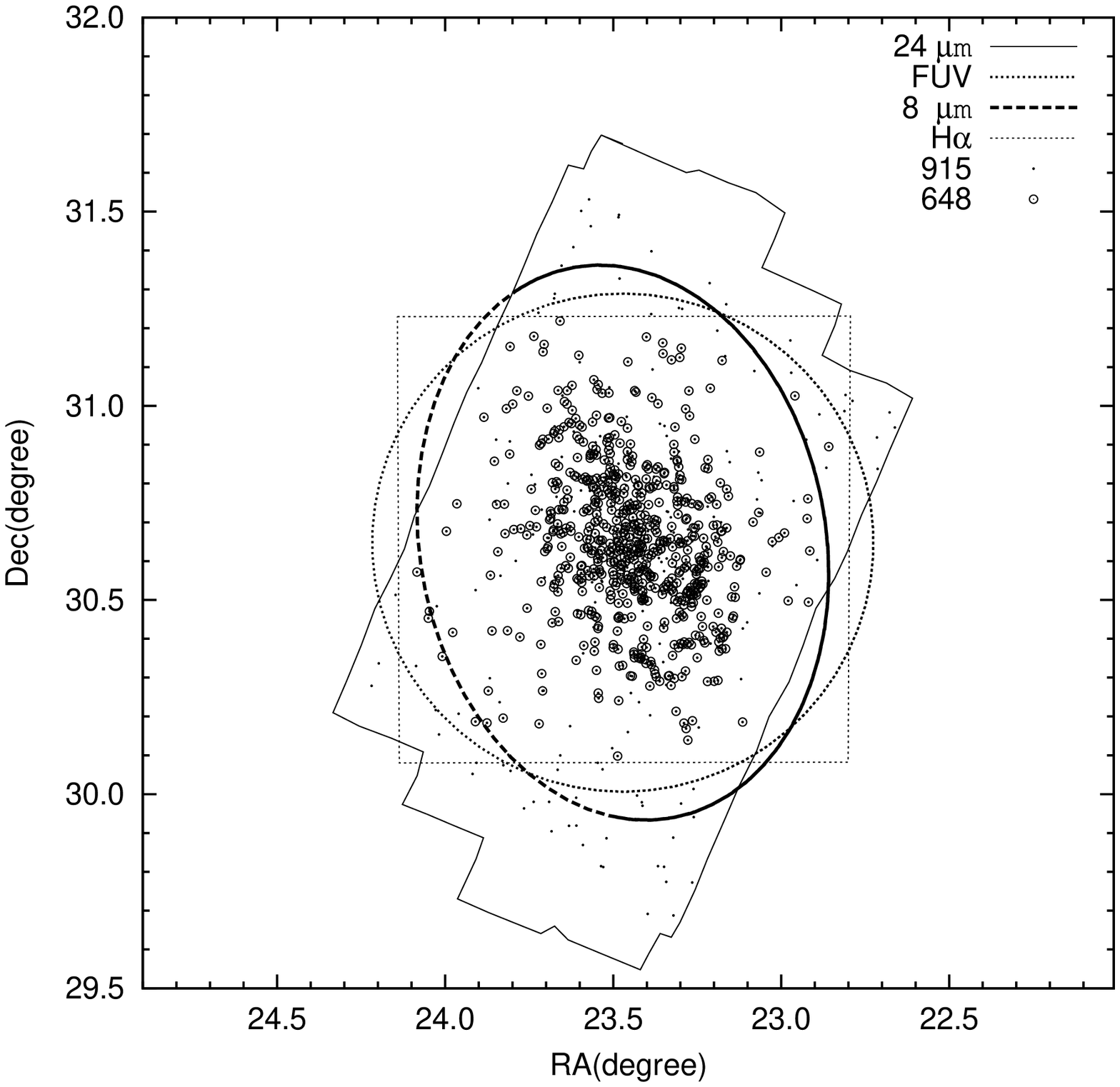}
\includegraphics[height=11cm,width=7cm,angle=-90]{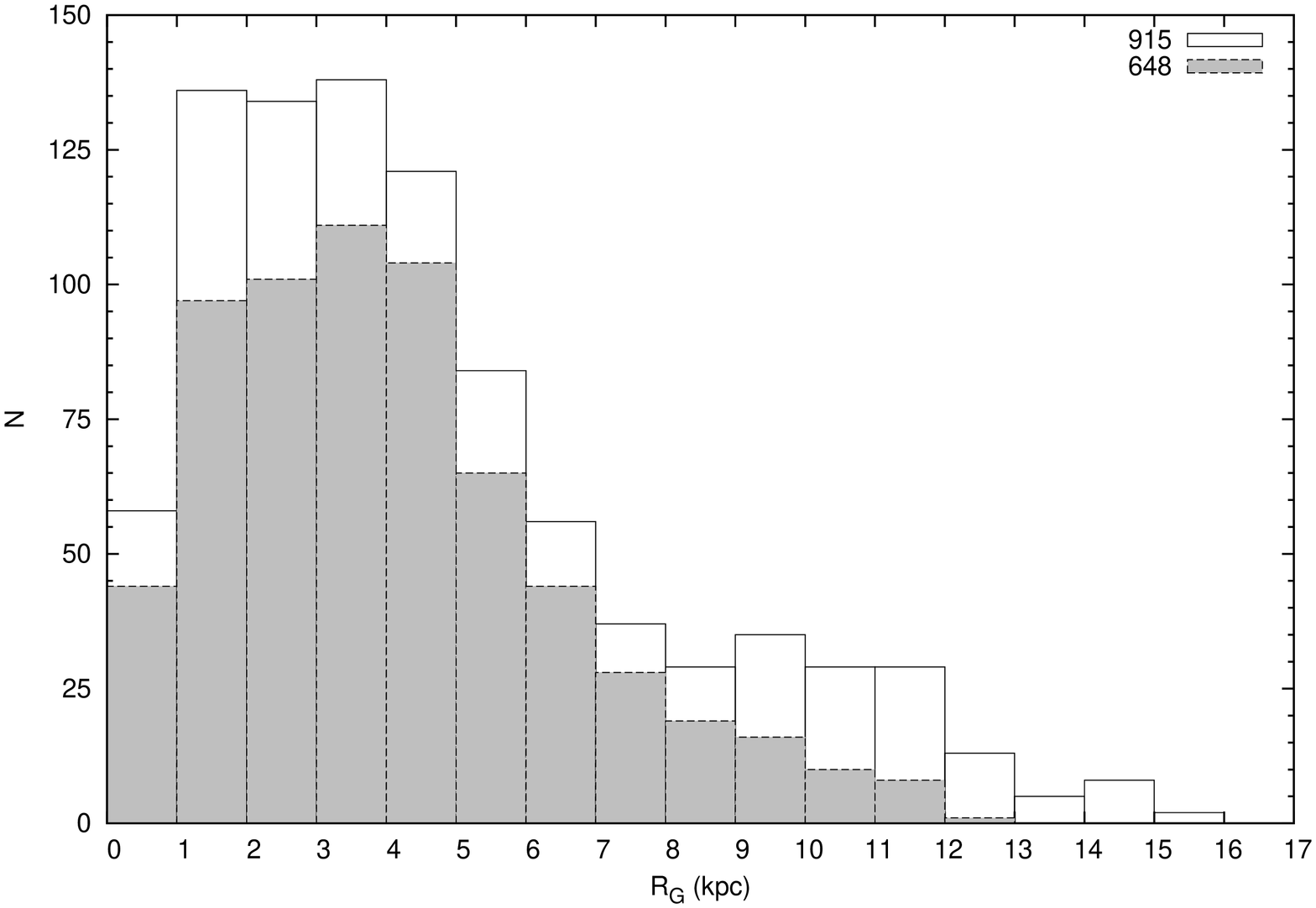}
}
\hbox{
\includegraphics[height=9cm,width=7cm,angle=-90]{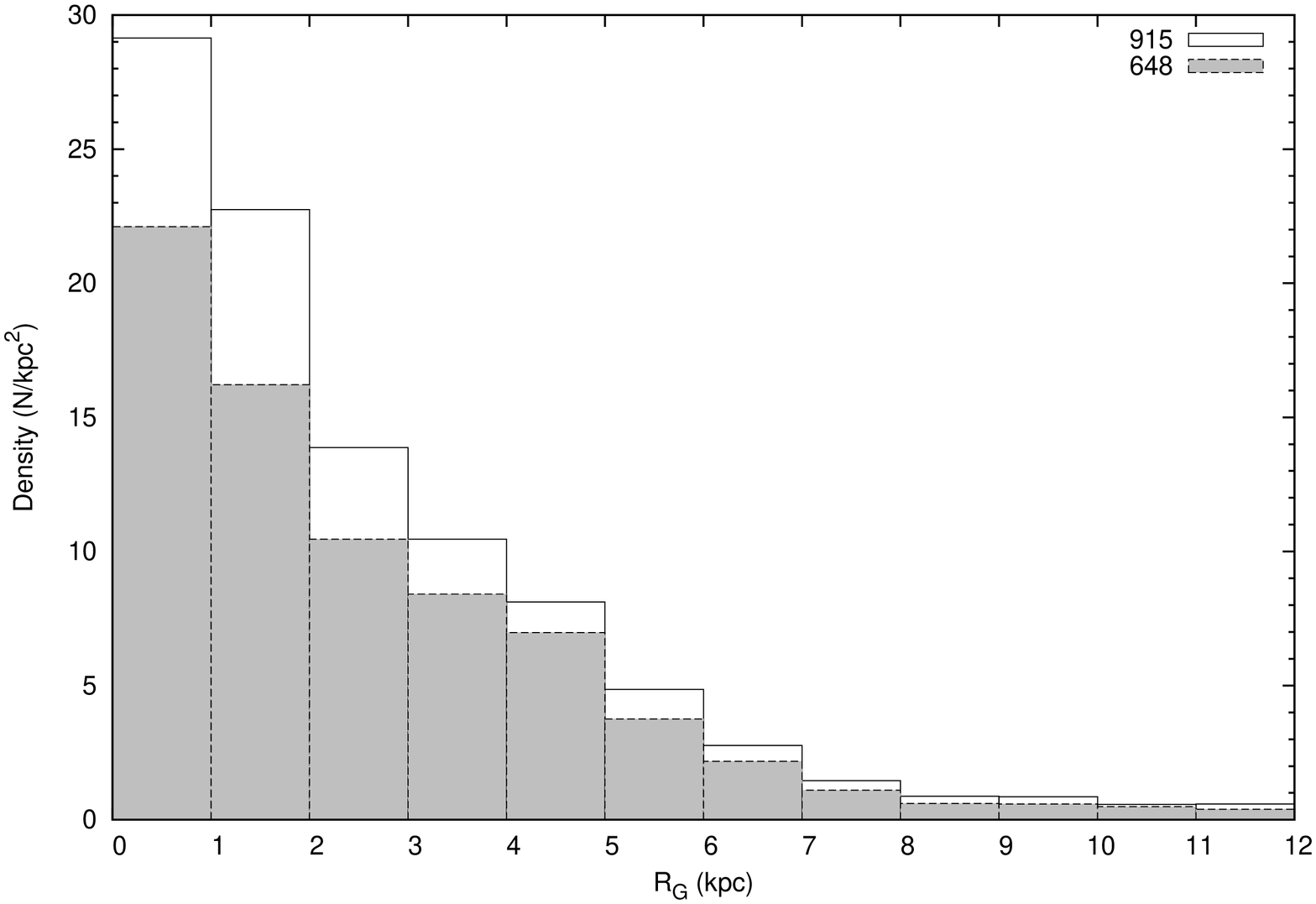}
\includegraphics[height=9cm,width=7cm,angle=-90]{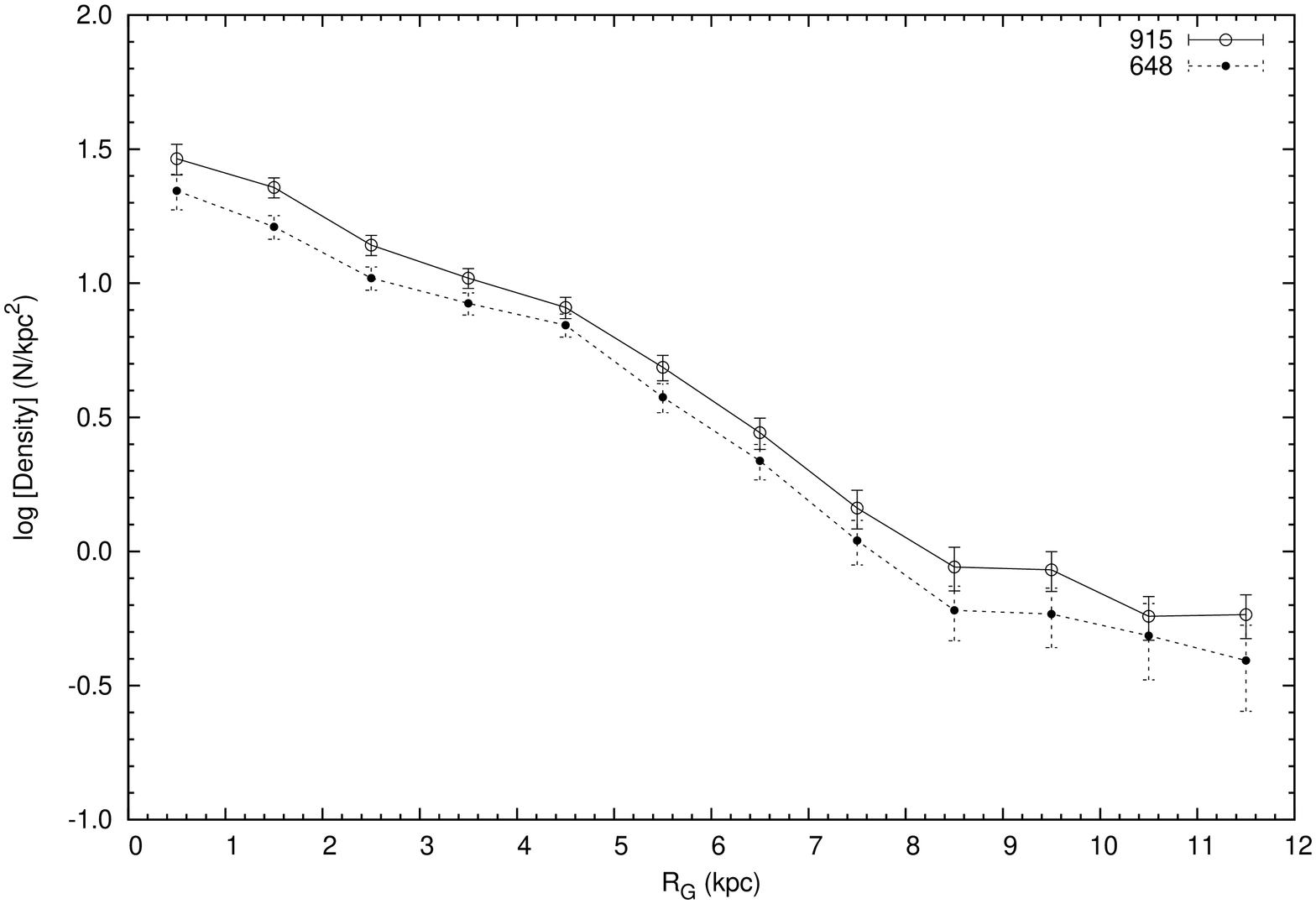}
}
\caption{\label{map} {\it (Top left)} Sky area covered by various maps along with the position of all the $24\,\mu$m sources (915, dot symbols) 
and the sample of YSCs (648, open circles). Coordinates are for the J2000 epoch. 
{\it (Top right)} Histogram of the number of clusters N as a function of the 
galactocentric distance $R_G$ for Mid-IR sources (white) and YSCs (grey). 
{\it Bottom left \& right panels} Profiles of the number density of clusters per unit area in kpc$^2$ in 
linear and log scale, respectively, as a function of galactocentric distance for the same samples. See text for details on the disk deprojection model.
}
\end{figure*}

The catalogue of YSCs of M33 includes all the Spitzer $24\,\mu$m sources with reliable photometry
in the FUV, NUV, and H$\alpha$. We are limited by the smaller extent of the GALEX and H$\alpha$
maps, compared to the mid-IR Spitzer maps, for the identification of $24\,\mu$m sources as YSCs. 
In the upper left panel of Fig. \ref{map} we have marked the area covered by the GALEX
maps together with the $24\,\mu$m mosaic and all the 915 sample objects.
The $24\,\mu$m map is  larger than that covered by GALEX and 
53 of the $24\,\mu$m sources lie outside of the latter. We were able to obtain reliable (error$<$0.5 mag) 
photometry for 789 sources in FUV, and 774 in NUV. A few other sources are
too weak in the maps for an estimate of the UV fluxes.
The adopted zero-points in the AB system \citep{1990AJ.....99.1621O} 
are 18.82 (FUV) and 20.08 (NUV).
We added, in quadrature, to the formal error provided by DAOPHOT
an uncertainty of 0.10~mag to allow for zero-point calibration errors
($\simeq 0.05$~mag) and large-scale background variance.

The total field of view of the H$\alpha$ map is $1.75\times 1.75$ deg$^2$
with a scale of 2 arcsec/pixel.
The H$\alpha$ map is also smaller than the $24\,\mu$m mosaic (see Fig. \ref{map} upper left panel)
and 55 of the $24\,\mu$m sources fall outside its borders.
We obtained reliable photometry (error$<$0.5 mag) for 772 sources.
We added in quadrature an error of 5\% in flux and corrected the H$\alpha$ flux assuming a 5\%
contamination by NII emission, as suggested by \citet{2000ApJ...541..597H}.

IR sources with no, or weak, H$\alpha$ counterpart are generally faint and their nature
is not obvious. A possibility is that they might be AGBs which are quite luminous in the mid IR.
Alternatively, they could be small sites of star formation
lacking massive stars, or deeply embedded clusters, or background objects.
\citet{2007ApJ...664..850M} have assembled a catalog of AGBs in M33
based on Spitzer data. We have cross-checked our list with this catalog
by searching for AGBs within the SExtractor $24\,\mu$m outer isophot.

In conclusion, out of the original 915 $24\,\mu$m sources, our final list of YSCs is made of 648 objects, 407 from the
catalogue of \citet{2007A&A...476.1161V} listed in Table \ref {T515}, and 241 from the newly selected sources 
given in Table \ref{T400}. The lack of the YSC label in the last column of Table \ref{T400} implies that
either the  MIR source is outside one of map boundary ( H$\alpha$, FUV or NUV) or it is a {\it bona fide} AGB  
(being associated to a MIR variable), or that it has unreliable photometry in H$\alpha$, or FUV- or NUV-band.
We retain as YSCs the few sources associated to MIR selected AGBs  which have a non-negligible number of ionizing photons,
log H$\alpha > 36.8$ erg sec$^{-1}$: these are large HII regions with an AGB star from a previous generation-star
or close by in projection.
 
The YSCs selected at $24\,\mu$m have a wide range in size and luminosity and they 
are found in a variety of environments: from fully isolated, mainly in the galaxy outskirts,
to crowded groups, especially in the inner spiral arms.
Sources in Table \ref{T400} are mostly in the outer regions and isolated.
In order to clearly assess the presence of UV and H$\alpha$ counterparts  
and the reliability of the extraction process, we made up and inspected an atlas where, for each source,
we assembled the images of the surrounding field in H$\alpha$, FUV, 8 and $24\,\mu$m.
This allowed us to select also a sample of  $24\,\mu$m sources with insignificant  FUV and H$\alpha$ brightness
as candidate embedded star forming regions which will be examined in a forthcoming paper. An insight on their nature 
could come from a possible association with a molecular cloud or clump \citep{2011A&A...528A.116C} but we have to
wait for deep enough CO maps with  sufficiently high spatial resolution to help in this regard.
 
\section{Radial densities and luminosity functions of IR selected sources}

Throughout the paper we assume a distance $D$ to M33 of 840~kpc 
\citep{1991ApJ...372..455F} to convert
fluxes into luminosities or to compute galactocentric distances.
Assuming that the sources lie in the disk and adopting the geometry and 
warp model of \citet{1997ApJ...479..244C},
described in detail in \citet{2000MNRAS.311..441C}, 
we derive the distance $R_G$ from the center of our sample objects.  
The radial distribution of the $24\,\mu$m sources and the selected YSCs
are plotted in the top right panel of Fig. \ref{map}. We can see that the MIR sources are found up to the largest distances,
while YSCs are distributed over the whole area of overlap of the various maps.

In Fig \ref{map} the bottom panels display the radial density 
distribution in linear and log scale, respectively.
The densities of the $24\,\mu$m sources and of the YSCs behave similarly
with an overall exponential decay with radius.
In particular, we note i) a shallow slope in the inner disk, 0-4.5 kpc;
ii) a steep decrease between 4.5 and 8.5 kpc; and iii) a flattening in the outer disk.
Past studies have found evidence for systematic radial variations in the star formation
history of the M33 disk. 
\citet{2007MNRAS.379.1302B} showed that the number of young stellar
groups drops suddenly at radii $>$ 4 kpc and attributed this result to a change in the properties of the star forming sites.
To study the inner and outer region clusters separately, we divide the $24\,\mu$m sources in `inner clusters' with $R_G<4\,$kpc
and `outer clusters' at larger distances. In this way, the size of the two groups is about the same.

\subsection{Estimate of the bolometric luminosity}

The bolometric luminosity of YSCs can be computed as the sum of the FUV and NUV luminosities
uncorrected for absorption, the H$\alpha$ luminosity corrected for extinction, 
$L^0_{H\alpha}$, and the total infrared luminosity $L_{TIR}$:

\begin{equation}
L_{bol}^{sum} = L_{FUV} +  L_{NUV} + L_{TIR} + 24L^0_{H\alpha}
\end{equation}

\noindent
where the terms H$\alpha$ account for the ionizing radiation
\citep{1999ApJS..123....3L} and $L_{TIR}$ for the UV radiation absorbed by
grains and re-emitted in the IR. We have not considered the 
radiation longward of 2800~A   which becomes important only for young clusters with luminosities lower than 
$10^{38}$~erg~s$^{-1}$. The luminosity in the FUV and NUV has been derived from

\begin{equation}
L_{UV} = \nu F_\nu 4 \pi D^2
\end{equation}

\noindent
with $D$ the distance to M33, $\nu = 1.95\times10^{15}$ Hz for the FUV and
$\nu = 1.3\times10^{15}$ Hz for the NUV, and

\begin{equation}
{F_\nu (Jy)} = 10^{23-(AB+48.6)/2.5}
\end{equation}

\noindent
AB indicates the corresponding magnitude in the FUV or NUV band.

\noindent
The total IR flux $(F_{TIR})$  and $L_{TIR}$ are computed from the expressions:

\begin{equation}
F_{TIR} = 10^{-14} \times (19.5 F_\nu(24) + 3.3 F_\nu(70) + 2.6 F_\nu(160) ),
\end{equation}

\begin{equation}
L_{TIR} =   F_{TIR}\times 10^3 \times  4 \pi D^2
\end{equation}

\noindent
where $F_{TIR}$ is in Wm$^{-2}$ and $F_\nu$ in Jy \citep{2002ApJ...576..159D}. 
The size of a large
fraction of sources in our YSC sample is below the Spitzer telescope resolution 
limit at 70 and 160~$\mu$m.
Hence, we cannot use the above expression to compute L$_{TIR}$ except for
a limited sample of extended resolved sources. We can estimate L$_{TIR}$ from the
emission at 8 and $24\,\mu$m if we find a good correlation between these and L$_{TIR}$.
For the sample of resolved sources at 70 and 160~$\mu$m we derive L$_{TIR}$ and
its relationship to the emission at 8 and $24\,\mu$m using the IRAC 8.0 $\mu$m 
photometry. This relation  has been
inferred previously but it might depend on the properties of the galaxy
considered and on the scale examined 
\citep[e.g.][]{2005ApJ...628L..29E,2006ApJ...648..987P}.
To find a relation between the 8 and $24\,\mu$m emission and $L_{TIR}$,
\citet{2009A&A...493..453V} plotted the $L_{24}/L_{TIR}$ ratio as
a function of the 8/24 $\mu$m flux ratio for their sample of stellar complexes 
in the inner disk and derived a linear fit:

\begin{equation}
log~L_{TIR} = log~L_{24} + 1.08 + 0.51~log({{F_\nu(8)}\over{ F_\nu(24)}})
\end{equation}

\noindent
where $L_{24}$ is $\nu F_\nu \times 4 \pi D^2 $.
We have checked that the above relation also holds in the outer disk i.e.
that the numerical coefficients of the above relation for sources in the outer disk are, 
within the errors, compatible with those relative to sources in the inner disk.
Thus, in this  study we adopt the relation of 
\citet{2009A&A...493..453V} to compute $L_{TIR}$ from the 8 and $24\,\mu$m fluxes. 
An error of 0.10 mag is added in quadrature to the photometric 
error to compute the uncertainties in the 8~$\mu$m flux. 
The  IRAC  $8\,\mu$m map is  smaller than the $24\,\mu$m mosaic
(see Fig. \ref{map} upper left panel) and 57 of the 915  sources fall outside it. 
For these sources and for the few (25) without a reliable 8.0~$\mu$m photometry (error$<$0.5 mag)
we  use the relation

\begin{equation}
log F_\nu (8) =  1.039\times log~ F_\nu (24) + 0.002
\end{equation}

\noindent
to derive the flux at 8$\mu$m which is needed to calculate $L_{TIR}$.
These coefficients are derived from a least square fit
to the sources detected in both bands.

Extinction corrections to the H$\alpha$ flux can be estimated using
$A_{H\alpha}=0.83 A_V$ with $A_V$

\begin{equation}
A_V = 1.7 \times log({{L_{TIR}\over {L_{NUV} + L_{FUV}}} + 1 })
\end{equation}

This is similar to the visual extinction formula used by \citet{2001PASP..113.1449C} 
if the ``fudge factor'' used to infer the luminosity outside the FUV band
is replaced with the luminosity in the NUV.

\begin{figure*}
\centering
\hbox{
\includegraphics[height=9cm,width=6cm,angle=-90]{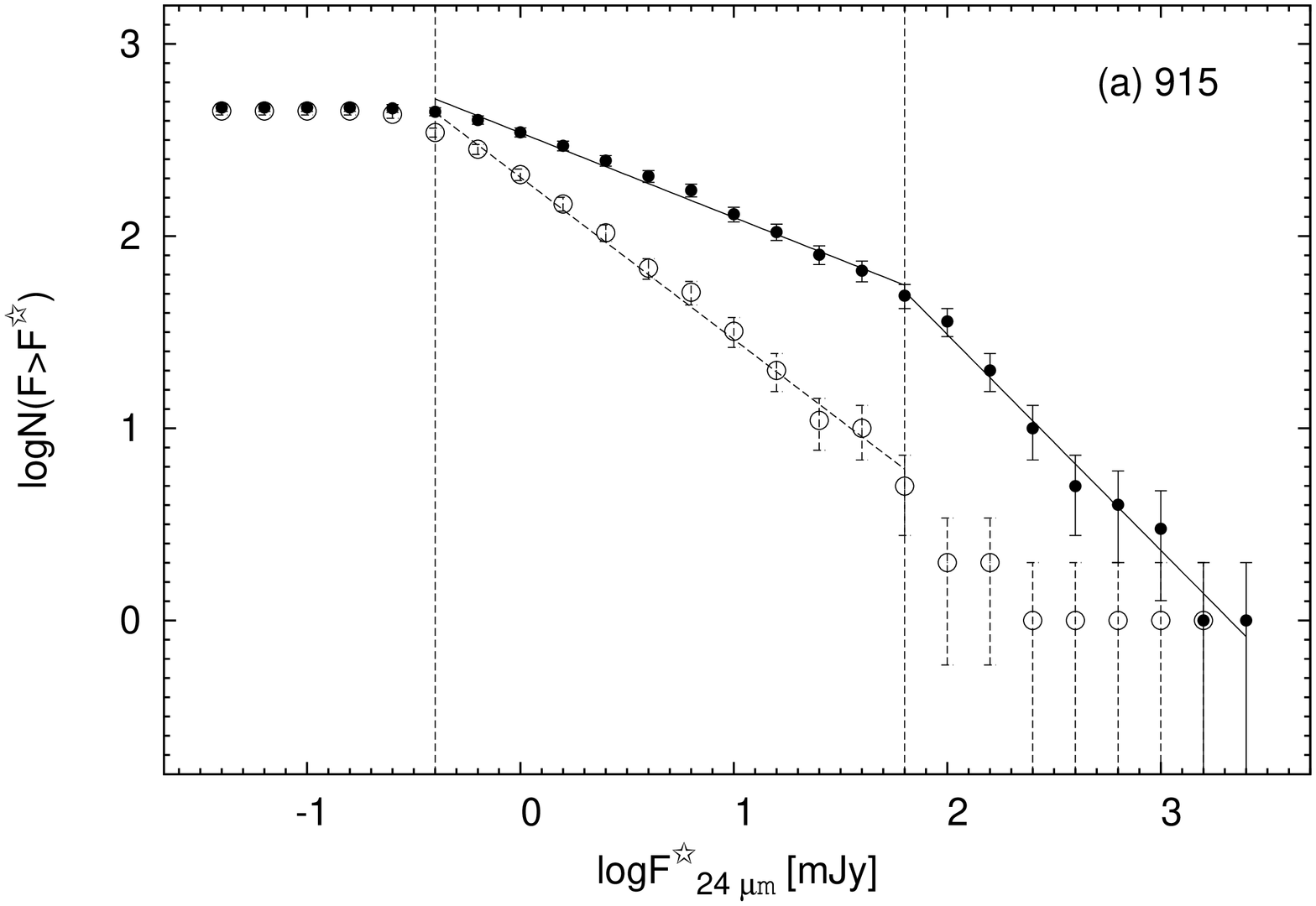}
\includegraphics[height=9cm,width=6cm,angle=-90]{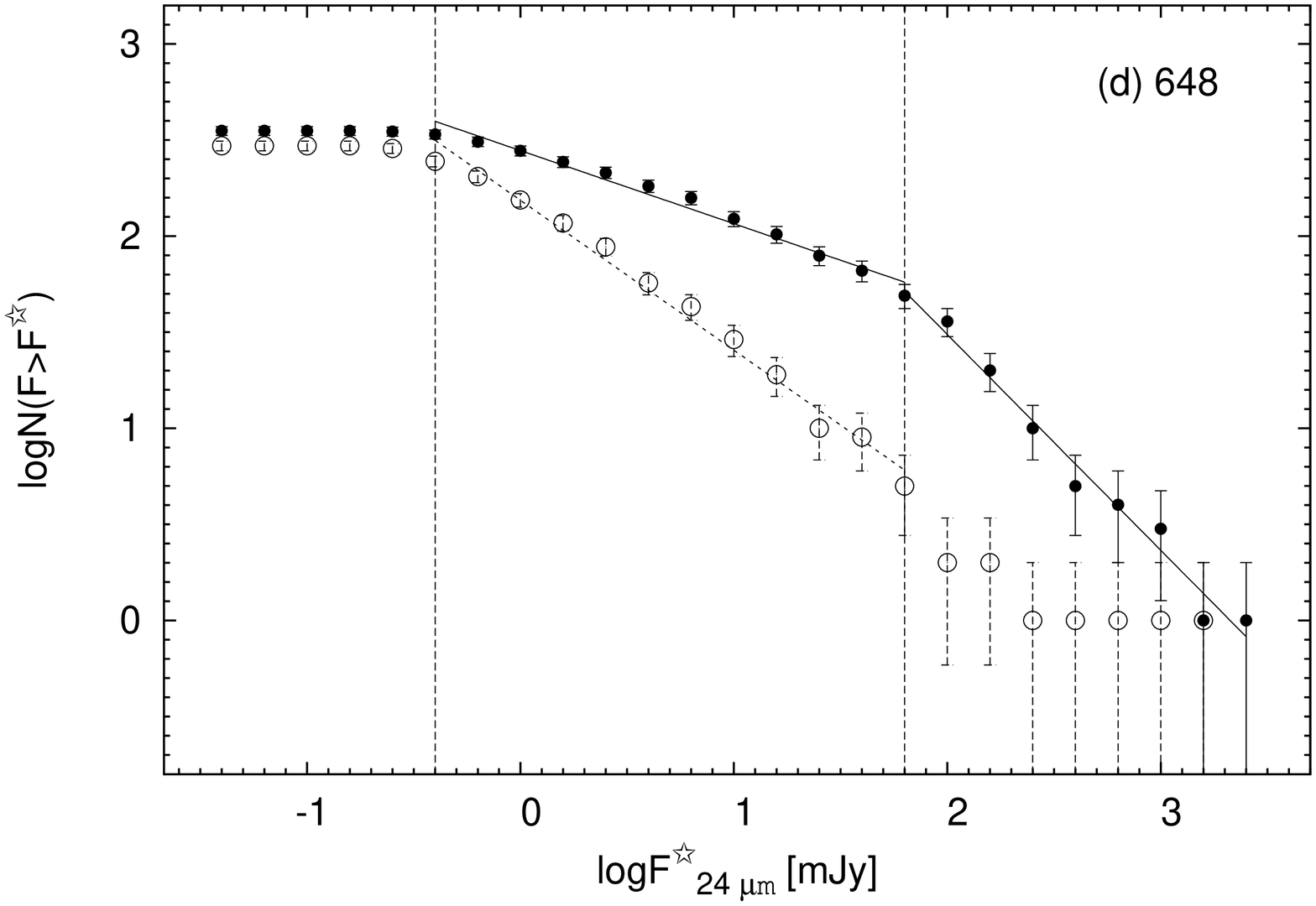}
}
\hbox{
\includegraphics[height=9cm,width=6cm,angle=-90]{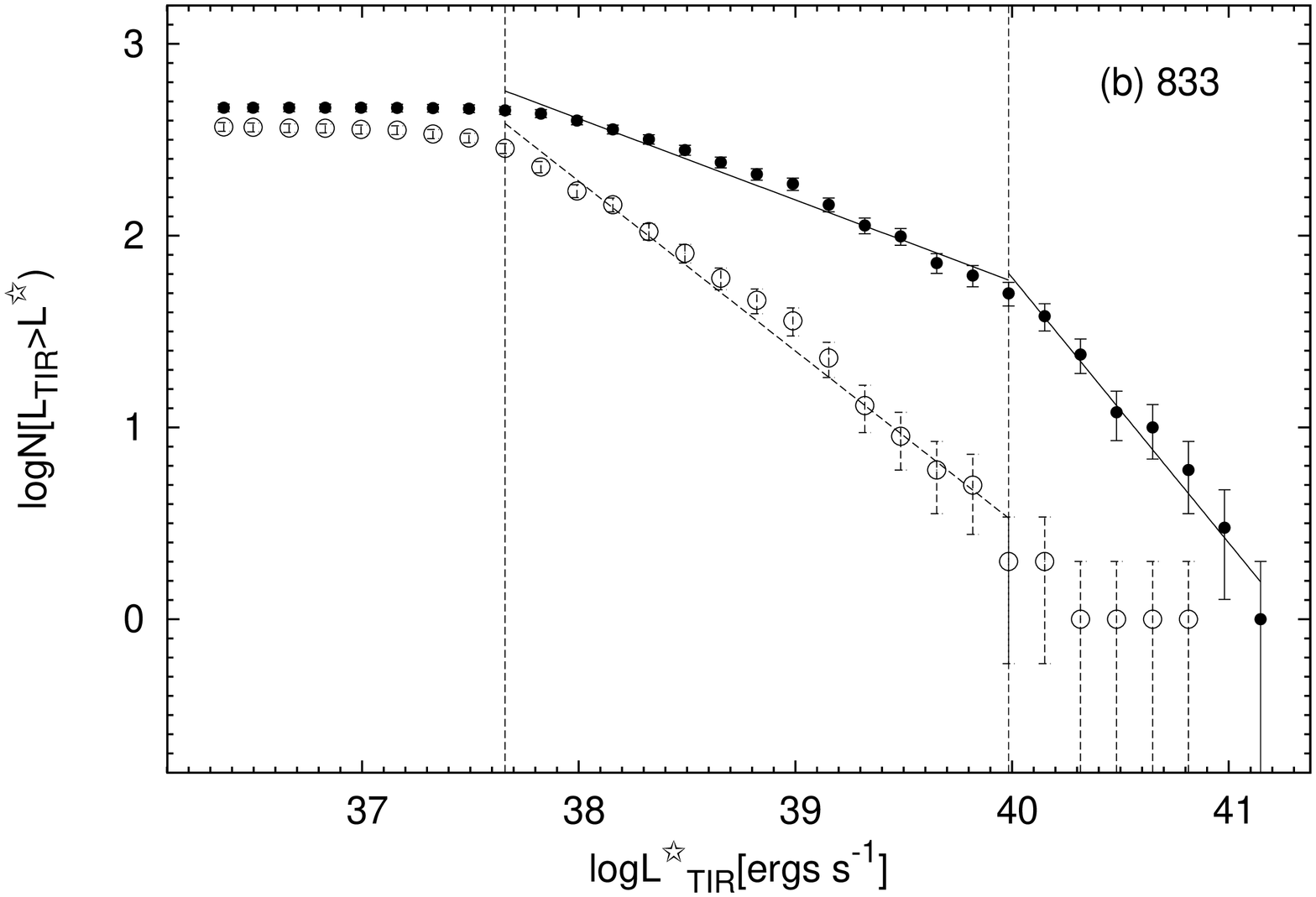}
\includegraphics[height=9cm,width=6cm,angle=-90]{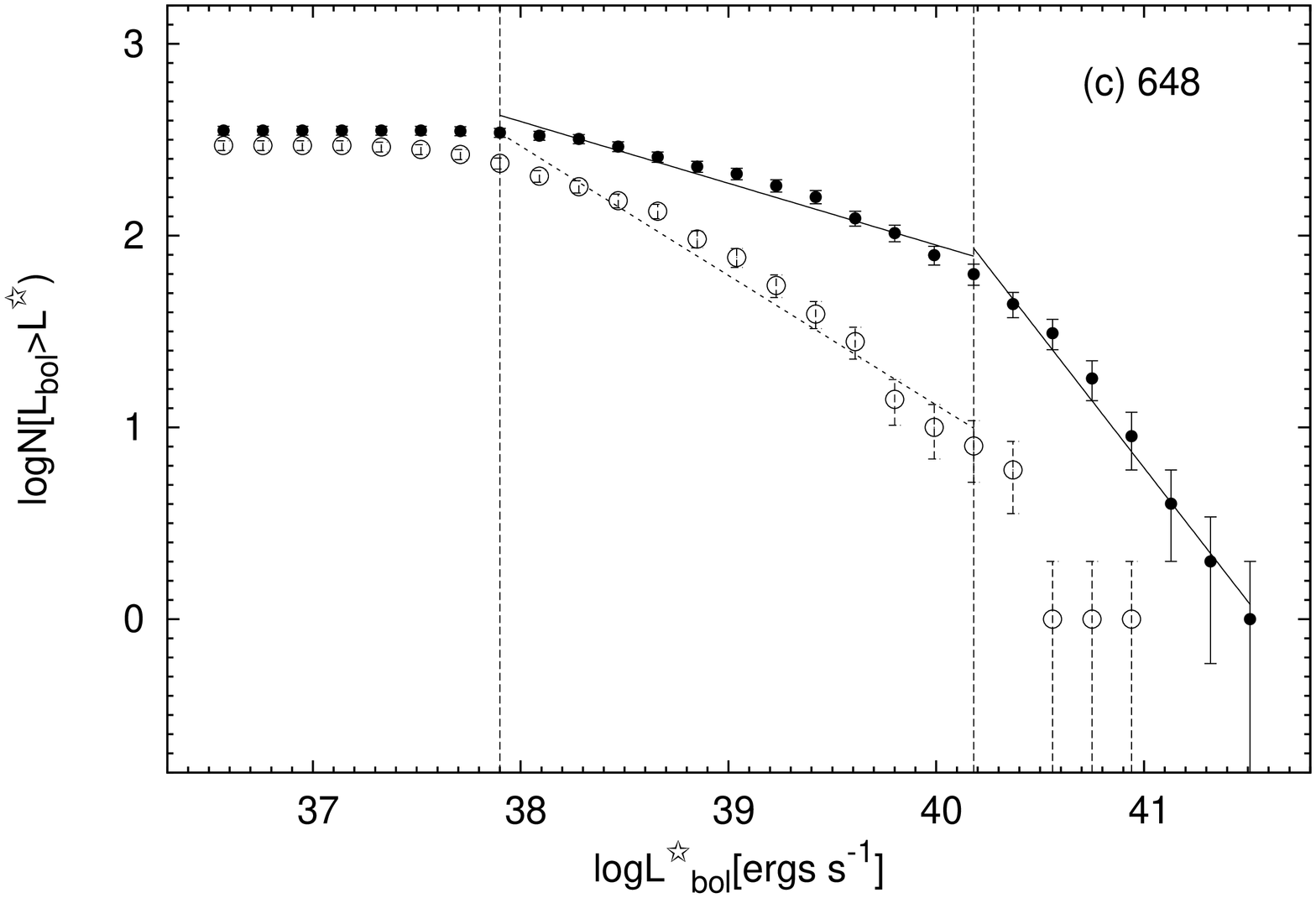}
}
\caption{\label{lf}The cumulative luminosity functions of: 
(a) the $24\,\mu$m sources (915), (b) the $8\,\mu$m sources (833) shown as a function of $L_{TIR}$, 
(c) the selected YSCs (648) as a function of $L_{bol}$ and (d) the selected YSCs (648), 
as a function of the $24\,\mu$m flux.
Filled and open circles represent counts in the inner and outer regions, respectively.
Linear fits for different flux ranges are shown as solid lines for the inner sample,
and dashed for the outer one.
}
\end{figure*}

\subsection{Luminosity functions}

The luminosity function (LF) of YSCs is a fundamental probe of the star formation process of a galaxy.
The cumulative LFs for the case of the inner and outer YSCs of M33 are shown in Fig. \ref{lf}.
We plot separately the results the cumulative function for the 24~$\mu$m flux for all the 915 $24\,\mu$m sources
(Fig.~\ref{lf}(a)), and the cumulative function for the TIR luminosity relative only to sources with an $8\,\mu$m 
counterpart (Fig.~\ref{lf}(b)). We see that the shape of the two LFs is similar; their
slopes are given in Table~3.  In the same figure we also plot the  cumulative LFs of the selected 648 YSCs
as a function of $L_{bol}$ (Fig.~\ref{lf}$(c)$) and of the $24\,\mu$m flux (Fig.~\ref{lf}$(d)$).
The values of the slope of the best least square fit to the data points between
$F_{\nu}(24\,\mu$m$) = 0.4$ to 70 mJy are listed in Table~3
as $1^{st}$ fit, while the $2^{nd}$ fit includes points with higher values of $F_{\nu}(24\,\mu$m$)$.
Since the values are lower than unity, the differential distribution $N(L)dL \propto L^{-\gamma}dL$
has $\gamma$ lower than 2.    

Overall, the main features of the LFs are the following:

(a) The slope of the cumulative function flattens at the faint end due to incompleteness. Crowding in the inner disk
implies a somewhat higher completeness limit.  

(b) The LF of the inner disk shows a definite change of slope for a TIR or Bolometric luminosity of order of 
10$^{40}$~erg~s$^{-1}$. The bright end slope is similar to that found
for bright HII regions in M33 \citep{1997PASP..109..927W} and in late-type spirals \citep{1991ApJ...370..526C}.

(c) The LF of the inner and outer YSCs differ markedly: there are very few bright outer 
sources and the LF is steeper at the faint end.  The presence of a break is not evident and, if present, it
occurs at a lower luminosity. The distribution can be well described by a single power law.

From the flattening of the distribution at the faint end, we
estimate the completeness limit of the catalog around 0.4 mJy
or $L_{TIR} \sim 5 \times 10^{37}$ erg s$^{-1}$. This corresponds to the bolometric luminosity of a single
B1.5V star \citep{2000asqu.book.....C}, indicating that our sample 
is close to being complete even for faint obscured HII regions.
The difference in the slopes of the inner and outer LFs
between $F_{\nu}(24\,\mu$m$) = 0.4$ to 70 mJy  
can be explained by a difference in the population of the star clusters. Namely,
the inner regions have more massive clusters with
the IMF fully populated up to the upper mass end. 
Conversely, the outer regions form predominantly clusters of lower mass
and hence the presence of massive stars is rare and stochastic, 
as we shall discuss in Section 5. Owing to the similarity between the LFs for the infrared and bolometric luminosity,
we can exclude the possibility that YSCs in the outer regions are fainter in the IR due to the formation
in an environment with a lower dust-to-gas mass ratio. As we will see below, the estimated YSC extinctions
show no radial variation. In fact, beyond the edge of active star formation the MIR-to-FUV or to-H$\alpha$ ratios 
increase. YSCs in the outer regions are intrinsically of lower luminosity: we
will discuss in the next Section whether this is due to an aging effect or to a different distribution in mass.

As for the inner clusters, the  distribution from the faint to the bright end can  be
broadly divided into two regimes with different slopes. The steep slope at high luminosity
is often observed in the LF  of HII regions,
open clusters and associations \citep{1997ApJ...476..144M}.
In the simple scenario, the change of slope represents
the transition from poor to rich clusters, where the latter are
populous enough to reproduce the high-mass IMF
\citep{1998AJ....115.1543O}. In this case, the transition point between
the two regimes marks the luminosity of the single brightest
star: below this value, the observed statistics is modified by
the sampling variance. We find that the transition occurs
around $F_{\nu}(24\,\mu$m$) \sim 60$~mJy,  i.e.  $10^{40}$~erg~s$^{-1}$, close to
the luminosity of an O3V star \citep{1996ApJ...460..914V}. 
This implies that most of the bright MIR sources are in fact luminous YSCs.

\section{SED fitting}

Using the photometry in the FUV, NUV, H$\alpha$ and MIR, we have constructed the SED of each YSC.
By a comparison with model SEDs from
Starburst99 \citep[][SB99]{1999ApJS..123....3L,2005ApJ...621..695V},
we can then derive the individual age, extinction, and mass;
in addition, we provide some estimate of the average properties of the
whole YSC population with implications on the IMF and on the ambient ISM.
Below, we illustrate the procedure.

\subsection{Stellar population modeling}

We model the YSC SEDs with the SB99 single-age stellar population synthesis. 
We consider that YSCs formed in an instantaneous burst
and assume a sub-solar metallicity of $Z = 0.004$ \citep{2010A&A...512A..63M}.
The IMF is chosen to be the default Kroupa IMF, that is
a broken power law ($\phi(M) \propto M^{-\alpha}$) with a slope
$\alpha=1.3$ for $0.1\,\msun < M < 0.5\,\msun$ and
$\alpha=2.3$ for $0.5\,\msun < M < M_{max}$ \citep{2001MNRAS.322..231K}.
Two sets of models were generated for different upper mass cutoffs:
$M_{max}=40\,\msun$ and  $M_{max}=100\,\msun$, respectively.
We use the Padova evolutionary tracks with
full AGB evolution to account for their
non negligible contribution to MIR fluxes. 
However, we did not consider the contribution of thermally pulsating AGB
stars that should be minimal for associations younger than
1 Gyr \citep{2006ApJ...652...85M}. The time evolution is followed with
logarithmic sampling (30 steps) from 1 to 100 Myr.

The SB99 calculations refer to a total stellar mass of $10^6\,\msun$,
and no allowance is made for dust extinction or emission.
The main output of the simulation is the emerging spectrum at each sampled age
which we then convolve with the GALEX and Spitzer filter bandpasses to
generate the synthetic photometry.
The model spectrum we use is the one taking into account both stellar and nebular emission,
the latter being computed for the case of ionization-bounded regions;
under this hypothesis, SB99 also provides the H$\alpha$ luminosity. 

First, in order to include the effect of extinction and dust emission in the SB99
photometry, we select a dust model among 3 cases made available by
\citet{2001ApJ...548..296W}\footnote{ http://www.astro.princeton.edu/~draine/dust/dustmix.html}:
Milky Way with $R_V=3.1$ (MW31), LMC ``average'' distribution {$\rm (LMC_{avg})$},
and LMC ``2'' distribution (LMC2).
For our purposes, the difference among them is mainly in the 
relative extinction between FUV and NUV. We first assume that A$_V$ is mainly produced in the 
neutral ISM (screen geometry).
Second, we choose a leakage factor $F_{lkg}$ as the fraction of the stellar radiation
that escapes unimpeded by interaction with the ISM, that is
without any extinction and re-emission in the IR and
also without ionization and H$\alpha$ production. We consider two values of the leakage factor of 0 and 1/3.
Third, we allow the possibility that a fraction $K_{dust}$ of the dust optical depth is
contributed within the ionized phase (dusty HII regions):  in this case, an important part of the ionizing 
flux is absorbed and reprocessed by dust, again without H$\alpha$ emission.
Again, we explore the influence of $K_{dust}$ for only two values,
0\% and 30\%, of the extinction within the HII region.

In conclusion, given a dust model, a value of $A_V$, the leakage fraction $F_{lkg}$, and the
fractional extinction in the HII region $K_{dust}$, we integrate the SB99 spectra and
determine, apart from a multiplicative factor, the corrected FUV, NUV,
H$\alpha$, and TIR luminosities, where TIR corresponds to the total
amount of radiation absorbed by dust.

\subsection{ $\chi ^2$ minimization:  Reddening, Age and Mass fitting}

\begin{figure*}
\centering
\includegraphics[height=10cm,width=16cm,angle=0]{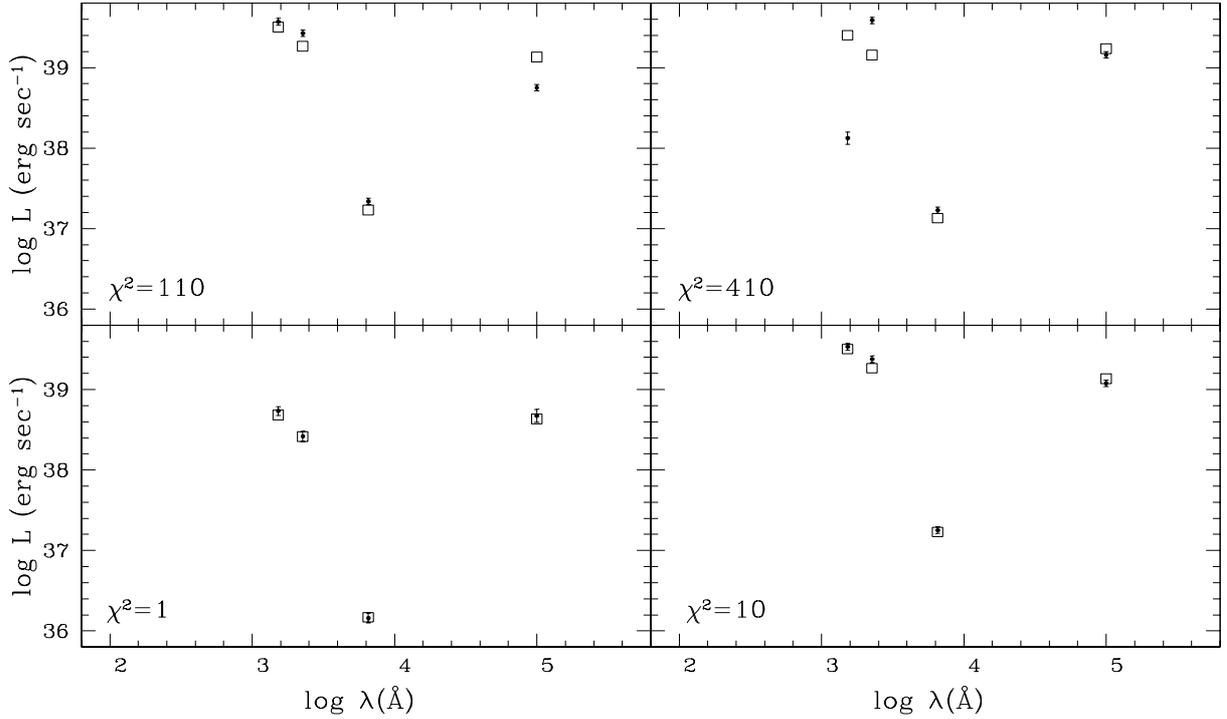}
\caption{\label{sed4} Data (filled dots) and best fit model (open squares) SEDs for four YSCs with small errorbars but
different $\chi^2$. Models in the
bottom panels give good fits to the data. Models with a $\chi^2$ of order 100, as in the top-left panel, are of 
lower quality, but still yield acceptable fits. The few sources whose best fit model has $\chi^2\gg 100$ have a SED inconsistent
with YSC models used in this paper.}
\end{figure*}

As outlined above, the corrected FUV, NUV, H$\alpha$ and TIR model luminosities
depend on 3 cluster variables (age  T, extinction A$_V$, and mass $M$)
and 4 population variables (upper mass cutoff, dust model, leakage, and 
HII region dustiness). For a given set of values of the population variables,
we used a $\chi^2$-minimization on each YSC and derived the best-fit
values of the cluster variables $(T, A_V, M)$

\begin{equation}
\chi^2(T,A_V,M) = \sum_N { [log(L_{obs}) - log(L_{model}) -A]^2\over \sigma^2_{obs}}
\end{equation}

\noindent
where $N$ is the number of bandpasses (4), $L_{obs}$ is the observed luminosity,  
$L_{model}$ is the model luminosity, and  $\sigma^2_{obs}$ 
is the estimated photometric error.
In practice, the age and extinction of each cluster are determined by comparing the shape 
of the SED to that of the SB99 models,
while the mass is the scale factor required to best match the absolute values
of the measured fluxes. Finally, $A = log(M/10^6\,\msun)$.
As an illustration of the method, we display in Fig. \ref{sed4} the SED best fits to four YSCs with different $\chi^2$ values.

\begin{figure}
\centering
\includegraphics[height=9cm,width=8cm,angle=-90]{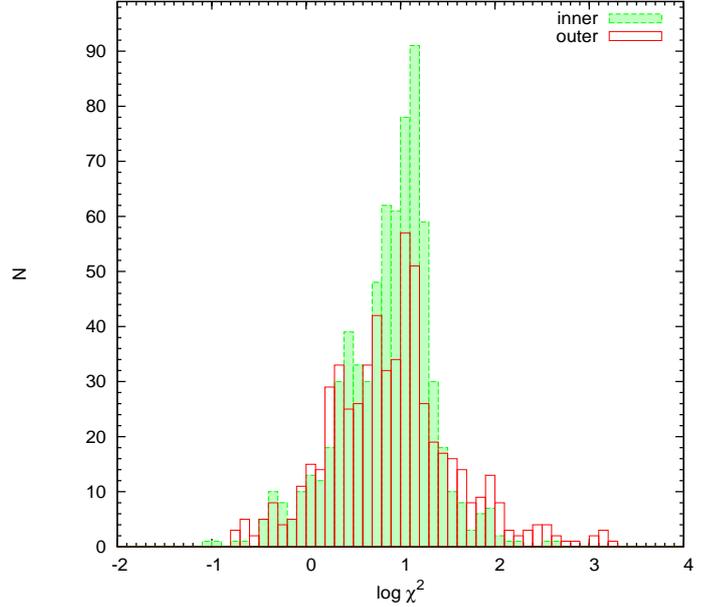}
\caption{\label{chiH} Distribution of the $\chi^2$ for the sample of inner and outer YSCs.  }
\end{figure}

The average $\langle\chi^2\rangle$ 
over the whole ensemble of YSCs of the minimized $\chi^2$'s is then used to
discriminate among the different combinations of population variables. 
Such comparison is also performed separately for inner and outer clusters.
Table \ref{Tsum}  lists the mean and median $\langle\chi^2\rangle$ 
for all the combinations of cluster variables. For the inner clusters we find that 
$\langle\chi^2\rangle$ has a minimum for 
$M_{max}=100, {\rm MW31}, F_{lkg}=0$, and $K_{dust}=0$, while for outer clusters the minimum is obtained 
for $M_{max}=100, {\rm LMC2}, F_{lkg}=0$, and $K_{dust}=0$.
The distribution of the $\chi^2$ for such combinations is shown in Fig. \ref{chiH}. 

We conclude
that the possibility of large-scale variations of the IMF, of radiation leakage, and
of dusty HII regions do not appear to be supported by these models that, however,
favor a systematic variation in the dust model.
The best fitting extinction curve for the inner
and outer disk in fact differs, varying  from Milky Way-type for the inner disk  
to LMC2-type for the outer disk.
The best  $\langle\chi^2\rangle$ is 14.8 for
YSCs in the inner disk and 42.7 for those in the outer disk. 
These values are rather large, despite the general goodness of the fit, because of
our conservative uncertainty estimate. 
Models with $\langle\chi^2\rangle$ values lower than 17.1 and 50.2  
for the inner and outer disk, respectively,
are within the 2$\sigma$ probability level, if we normalize the $\chi^2$ values to
$\langle\chi^2\rangle=14.8$. Thus, they cannot be 
ruled out. This implies that some dust
absorption inside the HII region is still a possibility,
but a 30\% or higher fraction of leakage of radiation is unlikely.

\begin{figure*}
\centering
\includegraphics[height=9cm,width=8cm,angle=-90]{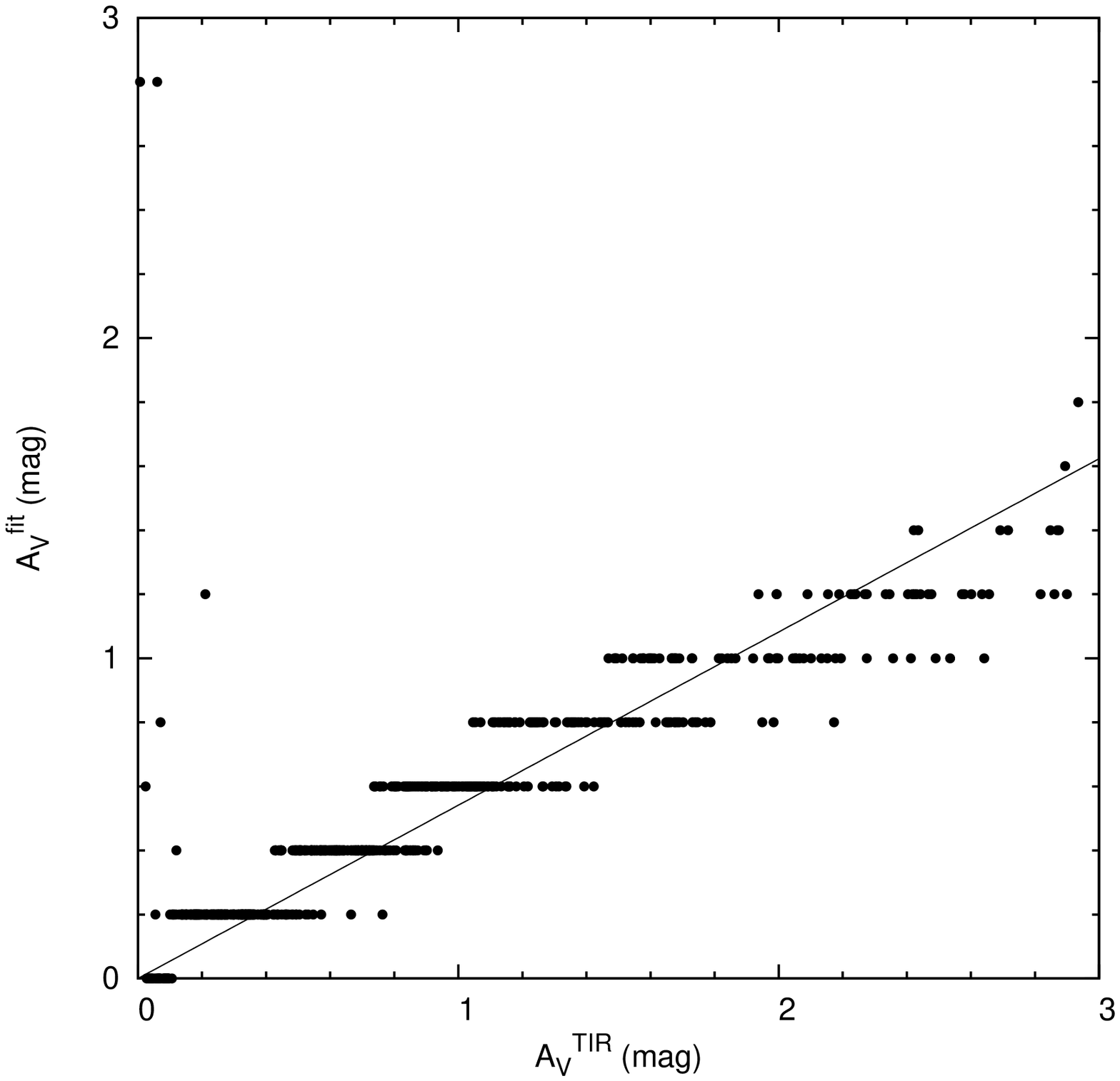}
\includegraphics[height=9cm,width=8cm,angle=-90]{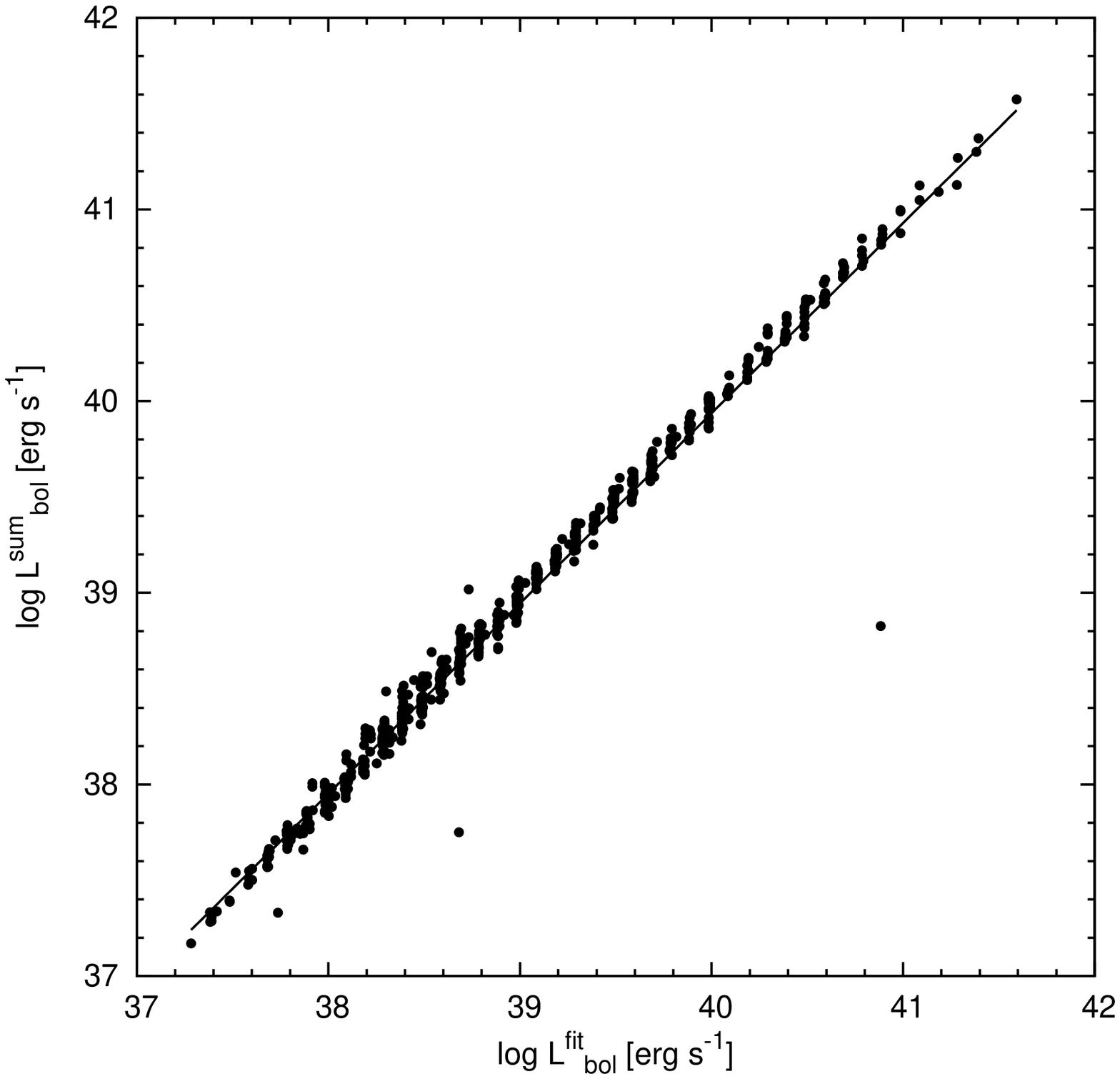}
\caption{\label{chi} {\it (Left panel)} Comparison of the $A_V$ values for the YSC sample derived from two different methods: 
$A_V$ as given in eq.(8) using the TIR luminosity ($A_V^{TIR}$), and using the SED model fits ($A_V^{fit}$). 
{\it (Right panel)} Same as the left panel, but for the luminosity estimated from the sum at different wavelengths ($L_{bol}^{sum}$) 
and from the SED model fits  ($L_{bol}^{fit}$). In both cases, the best linear fit is shown by the solid line.
}
\end{figure*}

For each YSC, we can then compare the estimates of $A_V$ and $L_{bol}$ 
from the available photometry (see eqs. (1) and (8)) 
to the values obtained with the $\chi^2$ minimization method. This is illustrated in Fig. \ref{chi}. 
While the $L_{bol}$ values are in excellent agreement,
the estimated $A_V$s using eq. (8) are
consistently higher than those inferred by the SED fits. The relation between the two quantities has a slope
of 0.54. Since eq. (8) is purely empirical and does not take into account 
the evolutionary state of the cluster and the dust model, we consider the fitted value as more reliable.

\begin{figure*}
\centering
\includegraphics[height=9cm,width=8cm,angle=-90]{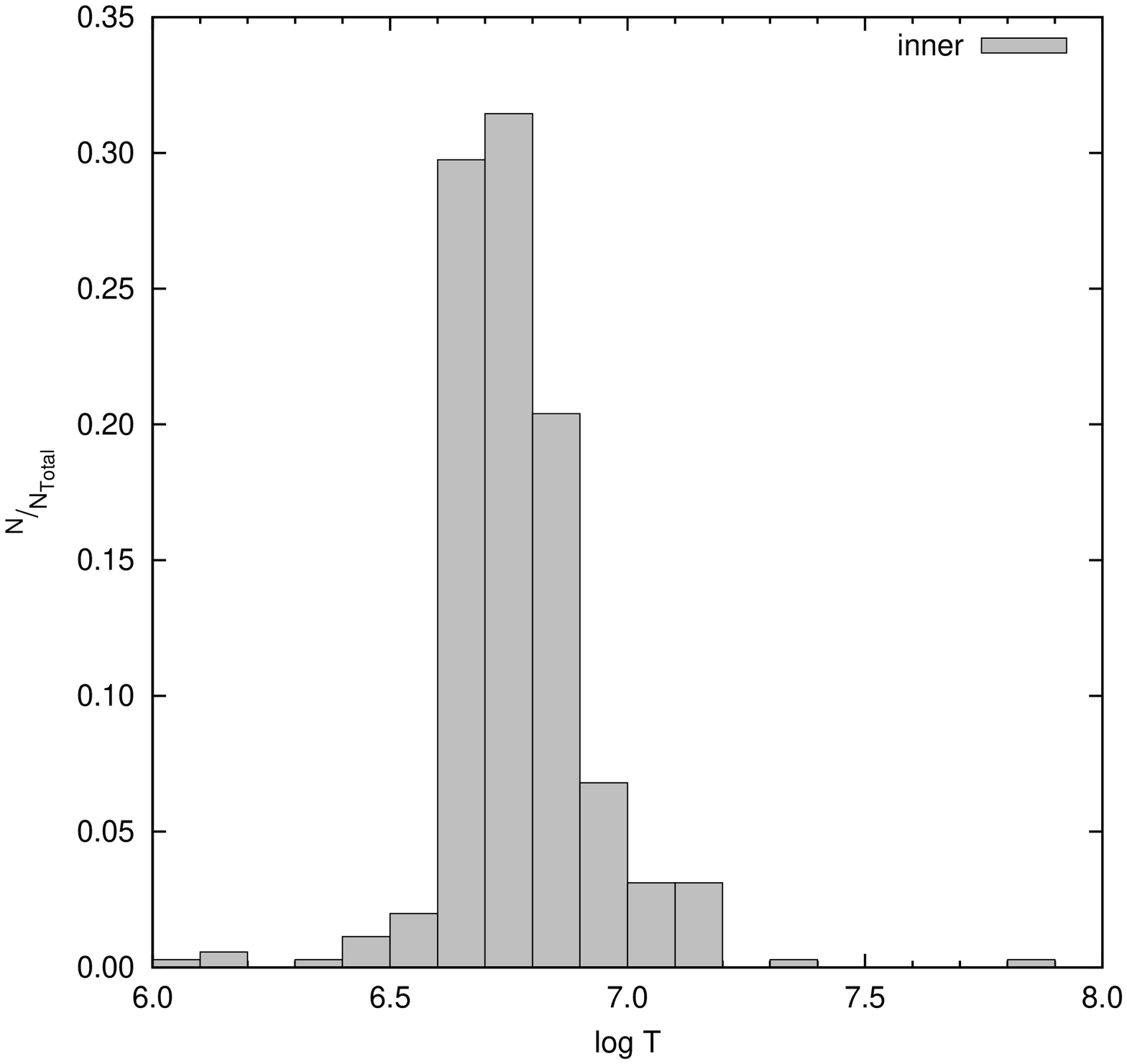}
\includegraphics[height=9cm,width=8cm,angle=-90]{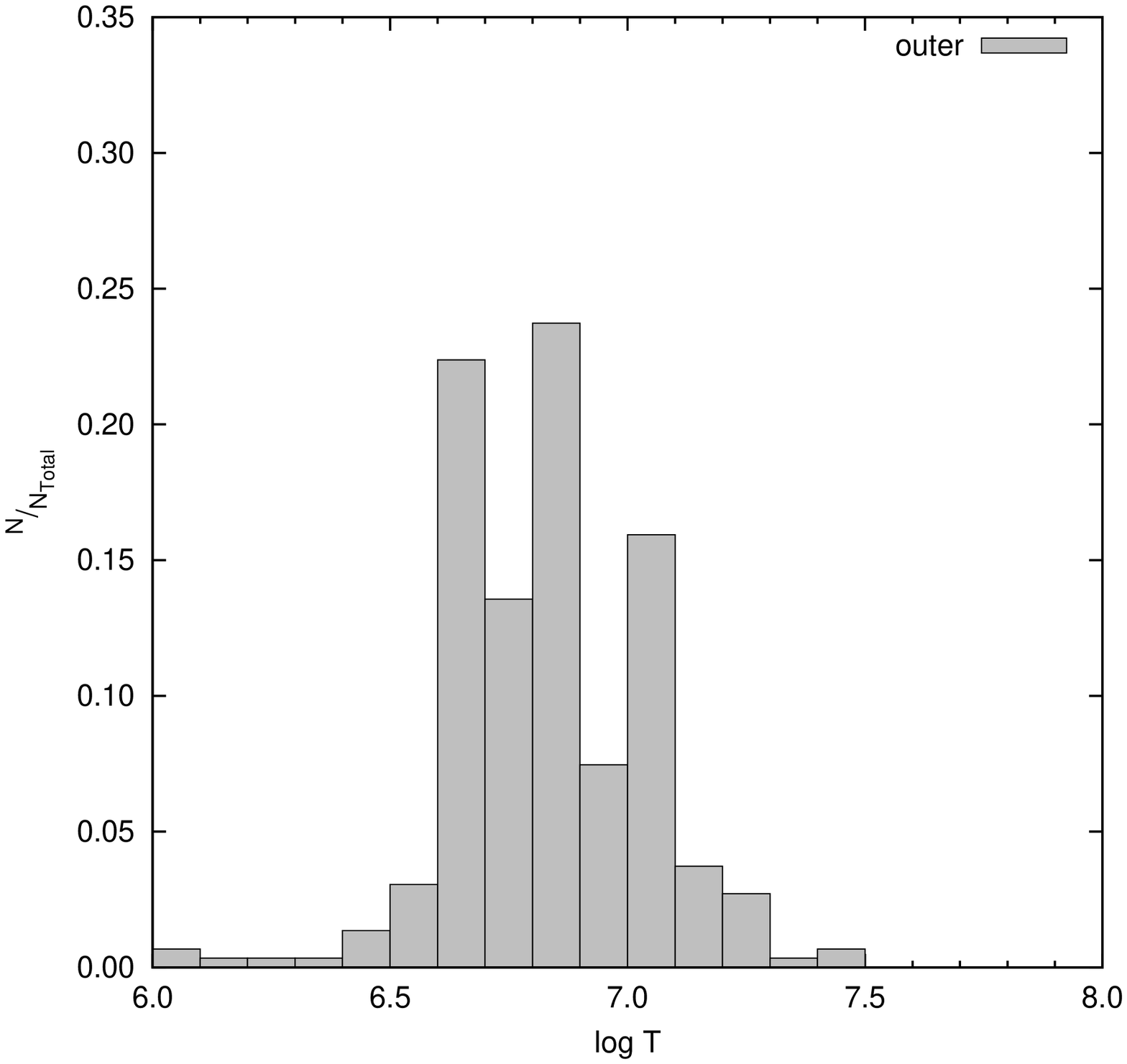}
\caption{\label{Hage}  Histogram showing the distribution of YSC ages in the inner and outer disk of M33.
On the x- axis  the age `T' is in a log scale; on the y- axis the number of clusters for each age bin is normalized to the total 
number of clusters in the inner (left panel) and outer (right panel) disk.
}
\end{figure*}

\begin{figure*}
\centering
\includegraphics[height=9cm,width=8cm,angle=-90]{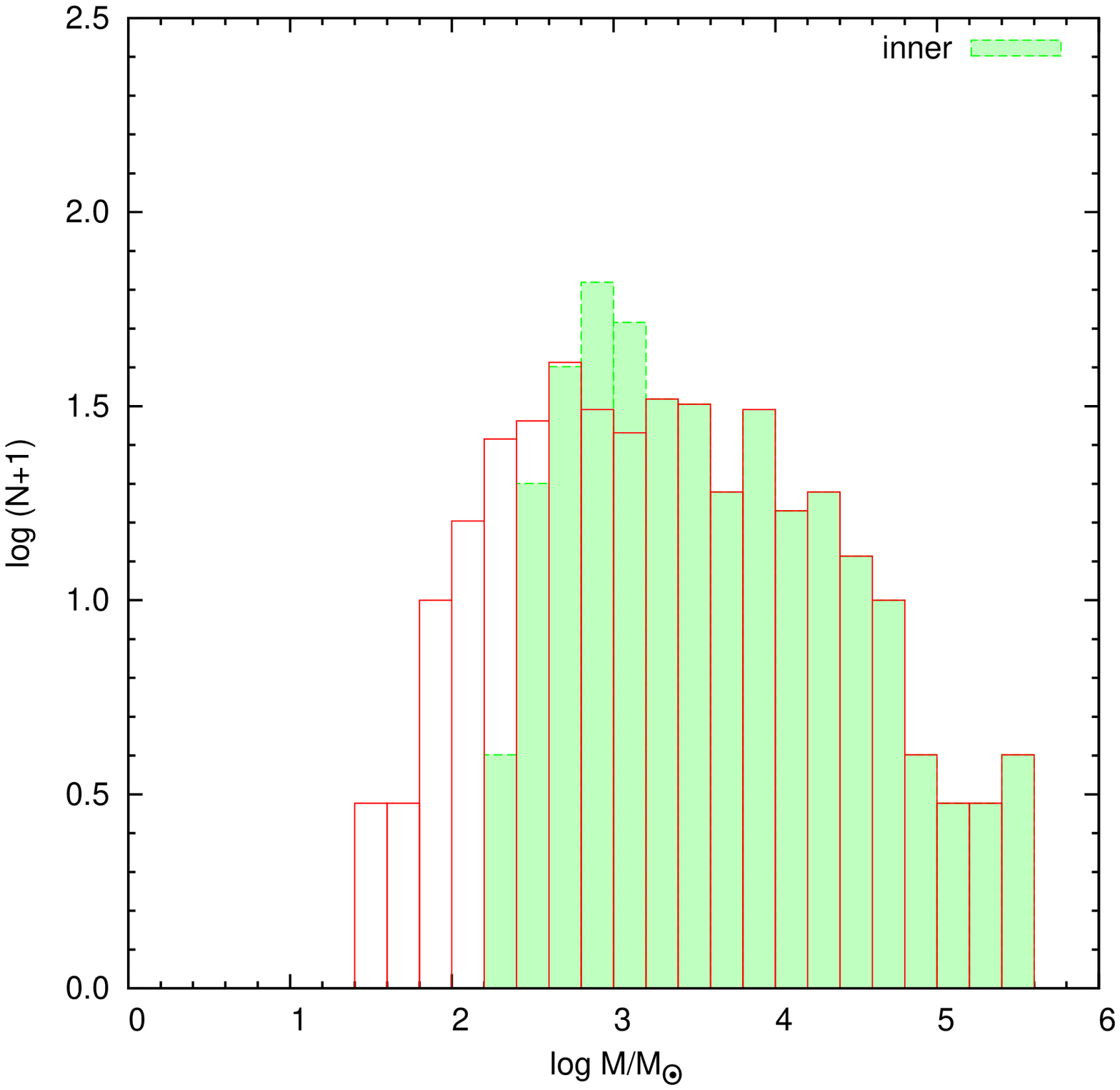}
\includegraphics[height=9cm,width=8cm,angle=-90]{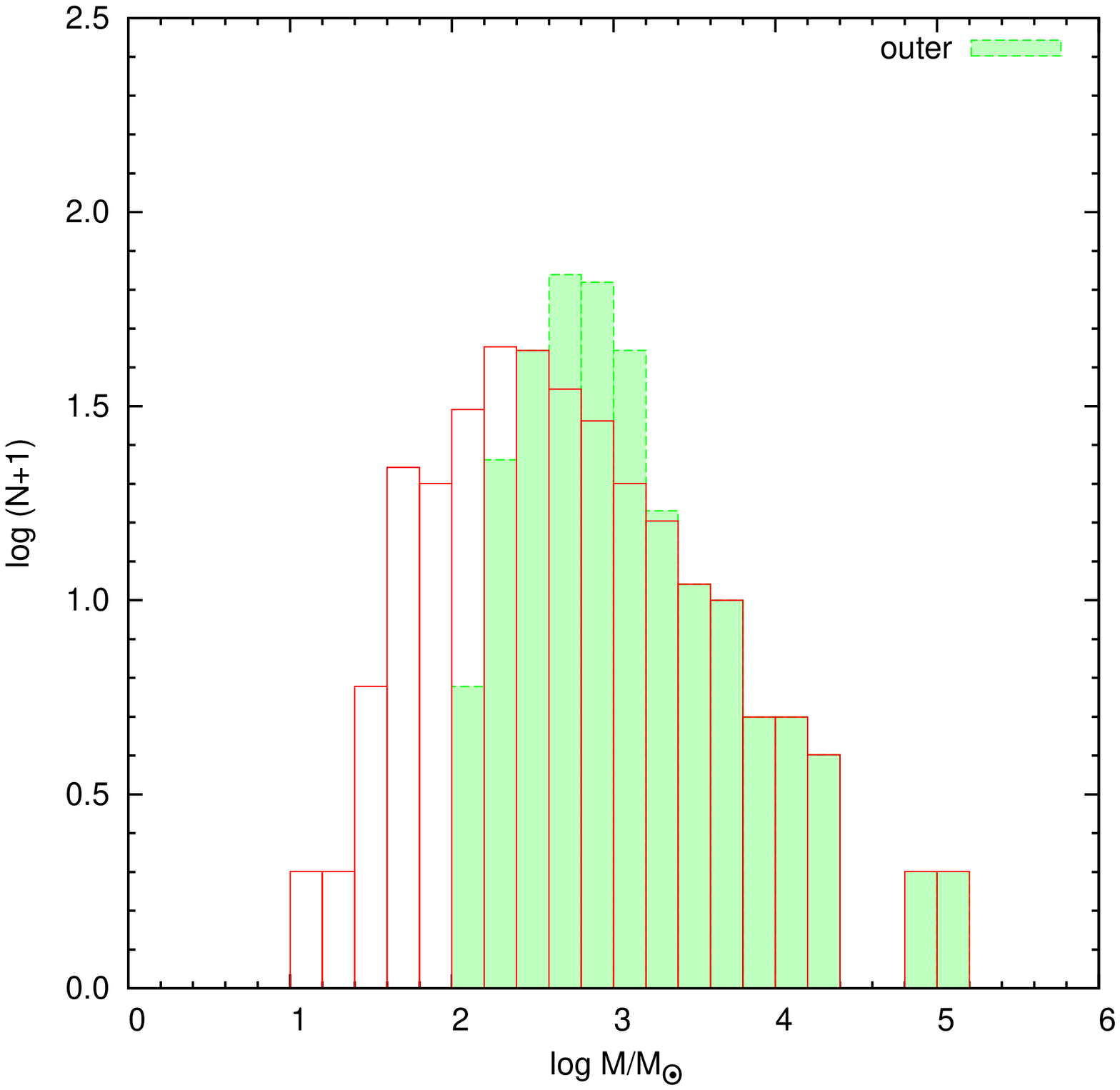}
\caption{\label{Hmass}  Distribution of the cluster mass in the inner and outer disk of M33.
We plot the log of mass (in solar units) and of N+1 (where N is the number of clusters in 
each mass bin), respectively.
The grey and white histograms are for the masses computed with or without incompletness corrections (see Section~5), 
respectively. 
}
\end{figure*}

The age, mass and extinction distribution of the inner and outer YSCs as derived from the best-fit SED models
are shown in Figure~\ref{Hage}, Figure~\ref{Hmass} and Figure{Hav} respectively.

\begin{figure*}
\centering
\includegraphics[height=9cm,width=8cm,angle=-90]{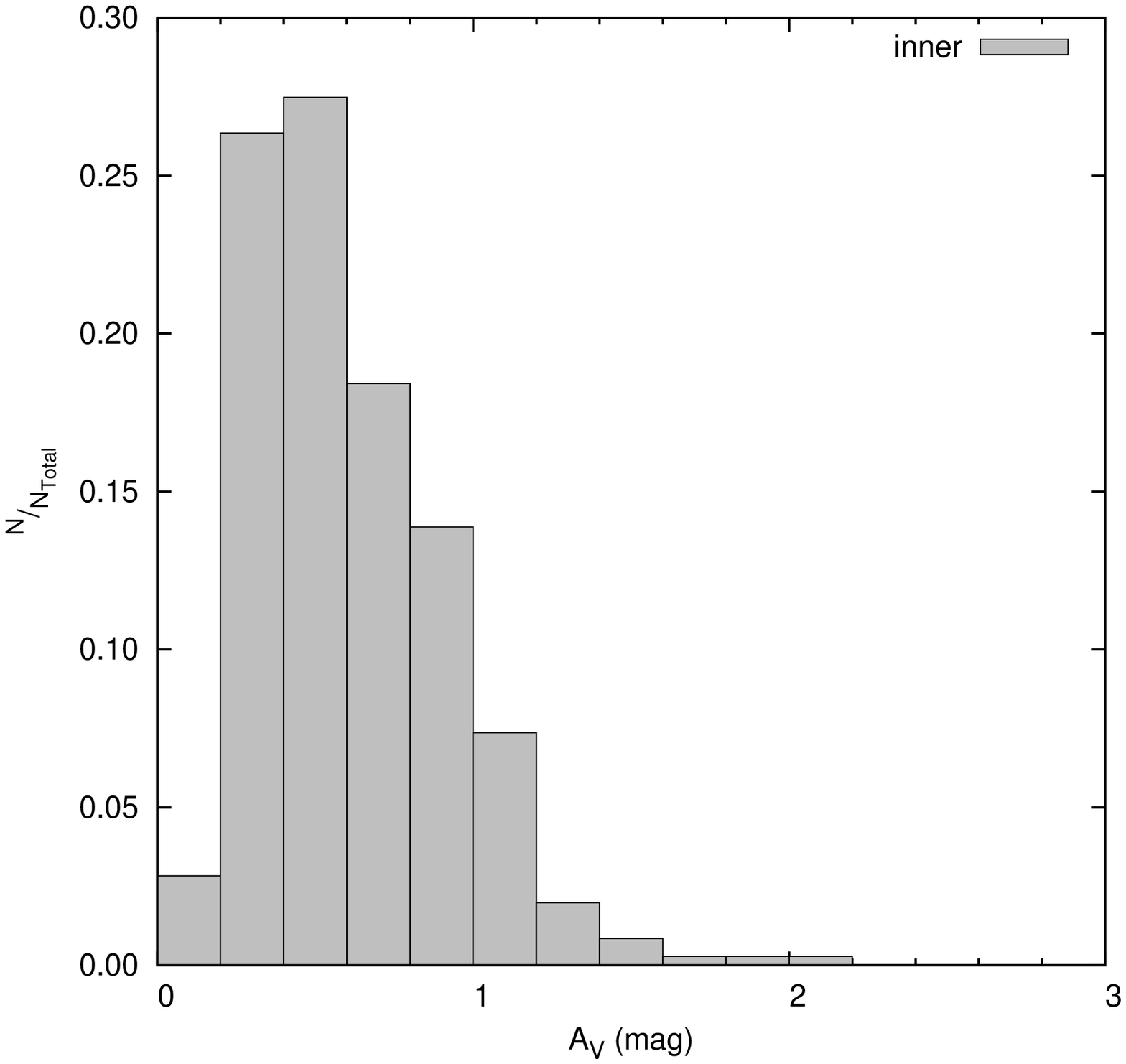}
\includegraphics[height=9cm,width=8cm,angle=-90]{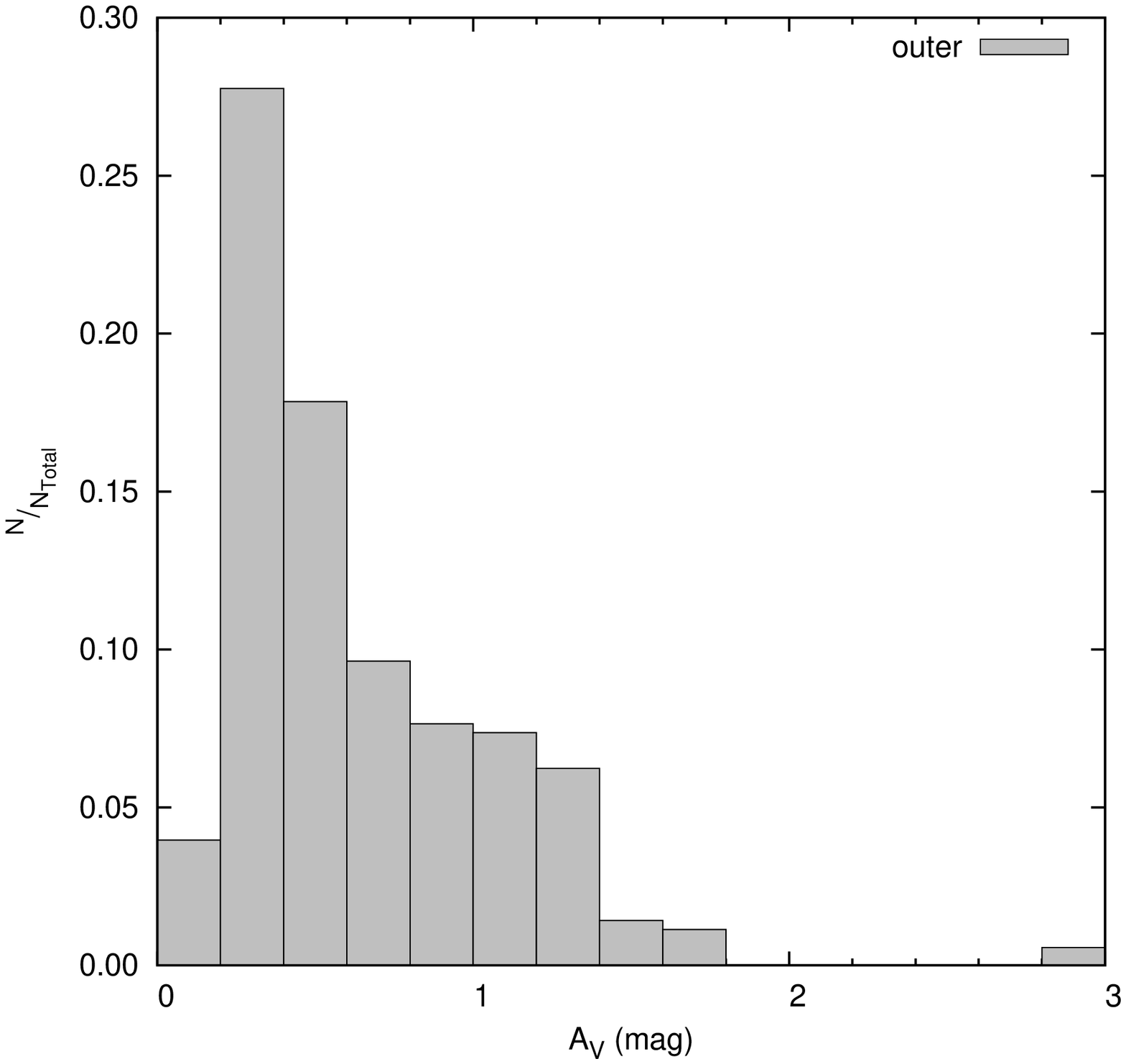}
\caption{\label{Hav}  Distribution of the visual extinction  in the inner and outer disk of M33.
$A_V$ is in magnitudes, while the number of clusters per bin is shown normalized 
to the total number in each sample.
}
\end{figure*}

\section{The massive stellar population of YSCs}

Among all the models employed in the fitting,
we find that the lowest average $\chi^2$ corresponds to a solution with the IMF populated up to
$100\,\msun$, with negligible leakage of radiation and dust absorption inside
the Str\"omgren sphere. 
From the $\langle\chi^2\rangle$ values given in Table \ref{Tsum} we can see that an IMF with a maximum mass 
lower than $100\,\msun$ has a lower probability but it cannot be ruled out. 
Models with an IMF fully populated only up to $40\,\msun$ have been used also to fit the SED of clusters with 
$L_{bol}<2\times 10^{39}$\,erg\,s$^{-1}$ only, or the sample of
clusters beyond 7~kpc (the nominal edge of the star-forming disk).
However, this choice did not provide an improvement
in the average $\langle\chi^2\rangle$. Therefore, we consider models with the upper mass cutoff at $100\,\msun$ 
as the best representation of the stellar population of YSCs throughout the M33 disk.

There is a clear difference between the mass distribution of clusters 
which form in the inner disk with respect to
those forming at large radii: beyond 4\,kpc clusters more massive than 
$10^4\,\msun$ are very rare. Similarly, beyond the H$\alpha$ edge at 7~kpc, 
most YSCs have masses smaller than $1000\,\msun$. 
If clusters form with total mass below $10^4\,\msun$, 
the number of stars is not high enough for the IMF to be fully populated
up to $100\,\msun$. 
It is well known that the most massive stars in a cluster tend to have a mass that increases
with the cluster mass \citep{1982MNRAS.200..159L}. 
This correlation leads to the question of
whether the upper limit to the stellar mass in a particular cluster depends on the
cluster mass because of some physically limiting process, or it is simply the result of
random sampling of the IMF.  In the first case, low-mass clusters cannot
produce high-mass stars \citep{2004MNRAS.348..187W,2006MNRAS.365.1333W}. 
In the second case they could as long as there is enough gas,
and intermediate-mass clusters should occasionally be found with unusually massive stars
-- ``outliers'' in the IMF \citep{2006ApJ...648..572E,2006A&A...451..475C}. 
A consequence of this model, often referred to
as stochastic or randomly sampled IMF model, is that
the summed IMF from many small mass YSCs  should be the same as the IMF
of a massive YSC. Considering the case for M33,
the higher $\langle\chi^2\rangle$ shown by the IMF
model with a mass cutoff lower than $100\,\msun$, 
even considering only the low-luminosity cluster sample or clusters beyond a certain radius, 
does not favor models which relate the maximum stellar mass to the cluster mass.
We recall that the results on the cluster birthline of M33  
support the stochastic model \citep{2009A&A...495..479C}, which also appears favoured by
the analysis of the IMF in more distant galaxies 
\citep{2010ApJ...719L.158C,2011arXiv1105.6101F}.
Hence, in what follows we will compare our data to the predictions of the stochastic IMF. 

We simulate the
cluster light and mass distributions considering a stochastic IMF 
from 0.1 to $100\,\msun$  with slope
$\alpha=-2.3$ for stellar masses greater than $0.1\,\msun$ and $\alpha=-1.3$ for lower masses.
We simulate 40,000 clusters that are distributed in number according to their mass between 20
and $10^4\,\msun$ as

\begin{equation}
N(M) dM =  K  M^\delta dM
\end{equation}

\noindent
where $K$ is a constant,
and $\delta$ is the spectral index of the Initial Cluster Mass Function, which we assume 
to be $\delta=-2$. 
This is in close agreement with the mass distribution of our IR selected YSCs 
(see also next Section) and with previous findings 
\citep[e.g.][and references therein]{2007ChJAA...7..155D}, even though the results of the simulations 
are not very sensitive to the value of $\delta$. 
We populate each cluster with stars randomly selected from the stellar IMF. 
More details on the numerical simulation are given in \citet{2009A&A...495..479C}.

Assuming the mean value $\langle\chi^2\rangle=14.8$ of the best fit model for the inner disk
YSC distribution as the  peak of the parent distribution, 
we consider YSCs whose best fit model has $\langle\chi^2\rangle\le 100$ well fitted by the
SED models. This is shown also by Figure~\ref{sed4}.  Leaving out clusters with $\chi^2>100$,
with large photometric errors (greater than 2 in log) and clusters older than
25\,Myr, we now compare the relation between the FUV and H$\alpha$ luminosities 
with that obtained by simulating a distribution of clusters at birth. 
We correct the observed FUV and H$\alpha$ luminosities for extinction using the
extinction value of the best fitting SED model. 

\begin{figure}
\centering
\includegraphics[height=12cm,width=9cm,angle=0]{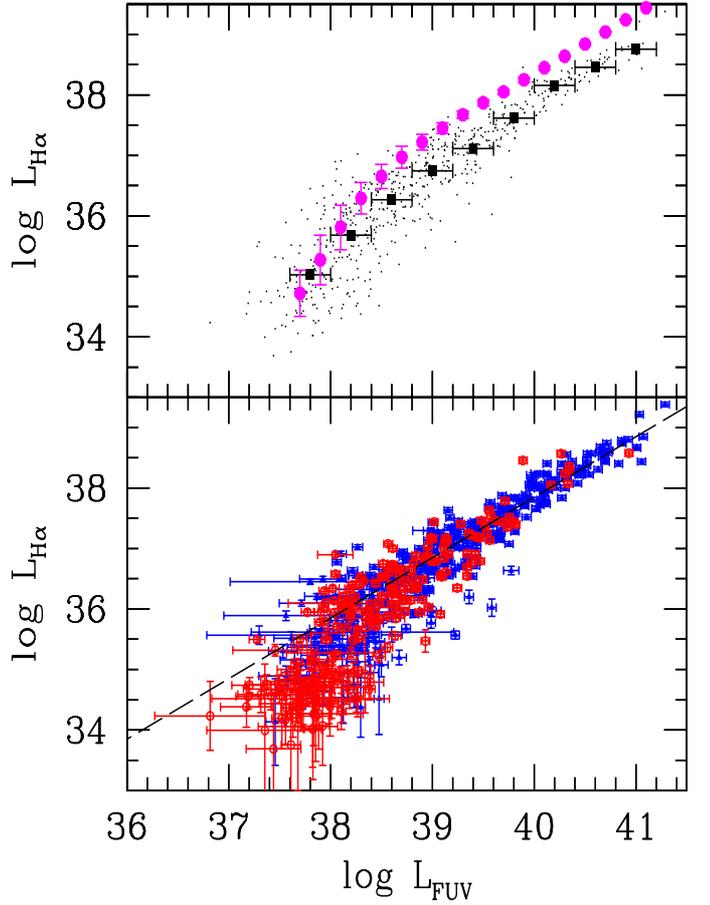}
\caption{\label{fuv} (Bottom panel) Distribution of H$\alpha$ luminosities 
as a function of FUV luminosities corrected  for extinction
according to the  best fitting SED model. 
Units are erg~s$^{-1}$.
The dashed line shows the linear relation consistent with the 
distribution for bright clusters. 
Filled triangles and open circles are for inner and 
outer disk YSCs, respectively. 
(Top panel) The same data are shown using small dots together with the average value
of $log L_{H\alpha}$ (filled squares) in 0.4 wide bins of $log L_{FUV}$. 
The filled circles show the 
expected values as derived from the simulation of the stochastic IMF for clusters at birth.
}
\end{figure}

In the bottom panel of Fig. \ref{fuv} we show the distribution of
H$\alpha$ luminosities as a function of the FUV luminosities corrected  for extinction
according to the best fitting SED model. 
The dashed line shows the linear relation which
is consistent with the distribution of the bright clusters. 
Clearly, the distribution deviates from the linear trend as  $L_{FUV}$ drops below
$10^{39}$\,erg\,s$^{-1}$.  
The sensitivity of our survey is good enough to trace the non-linear relation for 
$L_{FUV}>5\times 10^{37}$\,erg\,s$^{-1}$.
In the upper panel of the figure
the same data are shown together with the average value
of log $L_{H\alpha}$ (filled squares) in 0.4 wide bins of $log L_{FUV}$. 
The filled circles show the same
expected values as derived from the simulation of the stochastic 
IMF for clusters at birth (i.e. younger
than 3\,Myr). The most relevant result is that the expected deviations from the
linear scaling for a random sampled IMF model agree with the data. 
The measured H$\alpha$ luminosities for bright clusters are about 0.2 dex
lower than those predicted by the stochastic model. 
This is because our sample does not contain very young massive 
stellar clusters, and aging for massive clusters lowers the expected ${H\alpha}$ luminosity even 
if T$<10$~Myrs. 
From the figure we also see that stochasticity implies a large spread in the H$\alpha$ luminosities for 
$L_{FUV}<10^{39}$\,erg\,s$^{-1}$.

\begin{figure}
\centering
\includegraphics[height=9cm,width=9cm,angle=0]{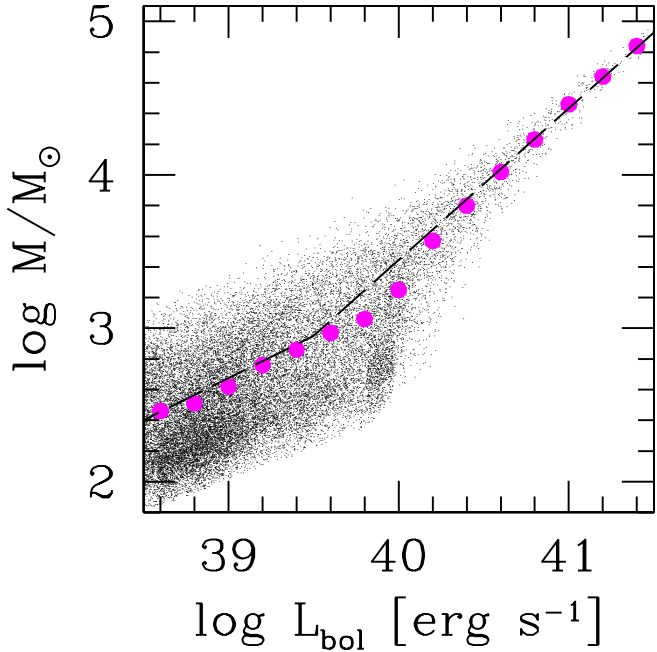}
\caption{\label{mb} Simulated distribution of cluster mass as a function of 
the cluster bolometric luminosities at birth. 
The large filled circles show the average cluster mass in
each bolometric luminosity bin. 
The broken dashed line is a fit to the average $log M$ -- $log L_{bol}$ relation:
at high luminosities it follows the linear relation between mass and luminosity
expected for a fully populated
IMF, while for log L$_{bol}<3\times 10^{39}$~erg~s$^{-1}$ it approximates 
well the simulation results.
}
\end{figure}

The presence of outliers in low-mass clusters populated according to a stochastic IMF
implies that the
relation between the cluster luminosity and its mass is no longer linear, 
or uniquely determined.
We show in Fig. \ref{mb} the distribution of cluster masses as a function 
of the cluster bolometric luminosity at birth. 
We have simulated the mass distribution using for massive clusters a $M$ -- $L_{bol}$
relation consistent with that given by the SED models for objects younger than 3\,Myr. 
The filled dots in the Fig. \ref{mb} are the average values of $log M$, 
in solar units,  for different values of log $L_{bol}$. 
For low-mass clusters, stochastic sampling of the IMF implies a mass higher, on average, than
that obtained from the extrapolation of a fully populated IMF. 
Deviations from a
linear scaling are evident as $L_{bol}$ approaches $10^{40}$\,erg\,s$^{-1}$ 
and become relevant for the average
mass when $L_{bol}\le 10^{39}$\,erg\,s$^{-1}$ .
A cluster of given luminosity can be of smaller
mass when some bright outlier dominates its integrated light, 
or can have a larger mass if no outliers are present.
The dashed line in the Fig. \ref{mb} shows the
linear approximation to the $log M$ -- $log L_{bol}$ relation for clusters at birth. 
We use this relation to derive the cluster mass from the fitted bolometric luminosity. 
Given a cluster age $T$, we define a critical luminosity

\begin{equation}
log L^{crit}_{bol}=47.3-1.2 log T
\end{equation}

\noindent
and, when $L_{bol} < L_{bol }^{crit}$, we correct for incompleteness using the expression:

\begin{equation}
log {M\over \msun}=0.58 log L_{bol}-24.45+0.7 log T .
\end{equation}

\noindent
In the above formulae $L_{bol}$ is in erg\,s$^{-1}$, 
$T$ is the cluster age in yr for clusters older than 
3~Myr and is $T=10^6$\,yr for younger clusters for
which the time evolution of $L_{bol}$ is negligible. 
For $L_{bol} > L_{bol}^{crit}$ the SED models show that
the cluster mass is well approximated by the relation 
$log M/\msun = log L_{bol}-44.3+1.2 log T$ .

Our main conclusion of the analysis of YSCs in M33 is that a stochastically sampled IMF with a maximum mass of $100\,\msun$
well describes many properties of their integrated light distribution throughout the whole disk.

\section{YSC properties}

We have corrected the mass of the less luminous 
clusters inferred by the SED model fits for incompleteness of the IMF according to the results of
the previous Section. The average mass of
low luminosity clusters is underestimated if no correction is applied.  

\begin{figure*}
\centering
\includegraphics[height=9cm,width=8cm,angle=-90]{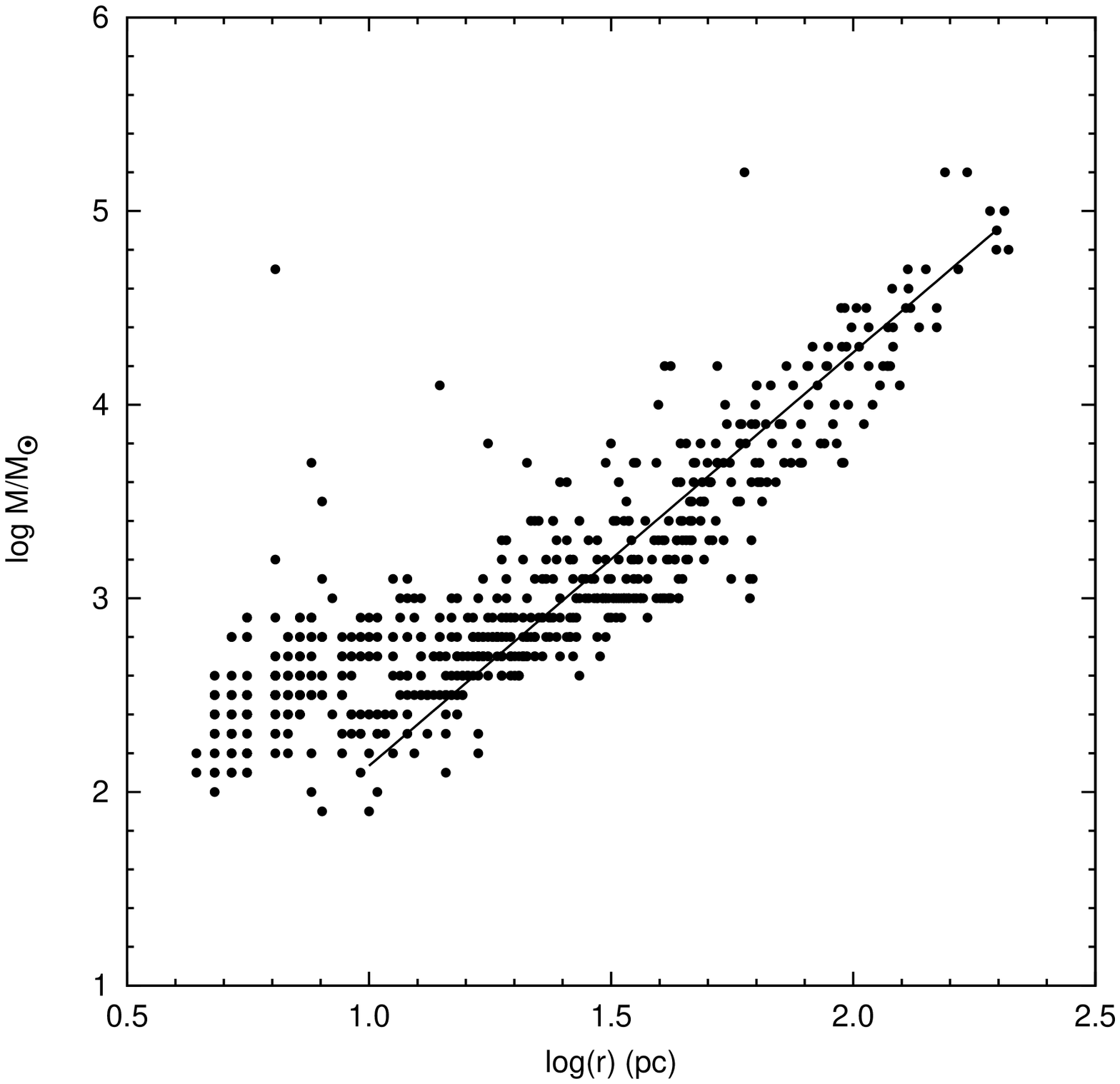}
\includegraphics[height=9cm,width=8cm,angle=-90]{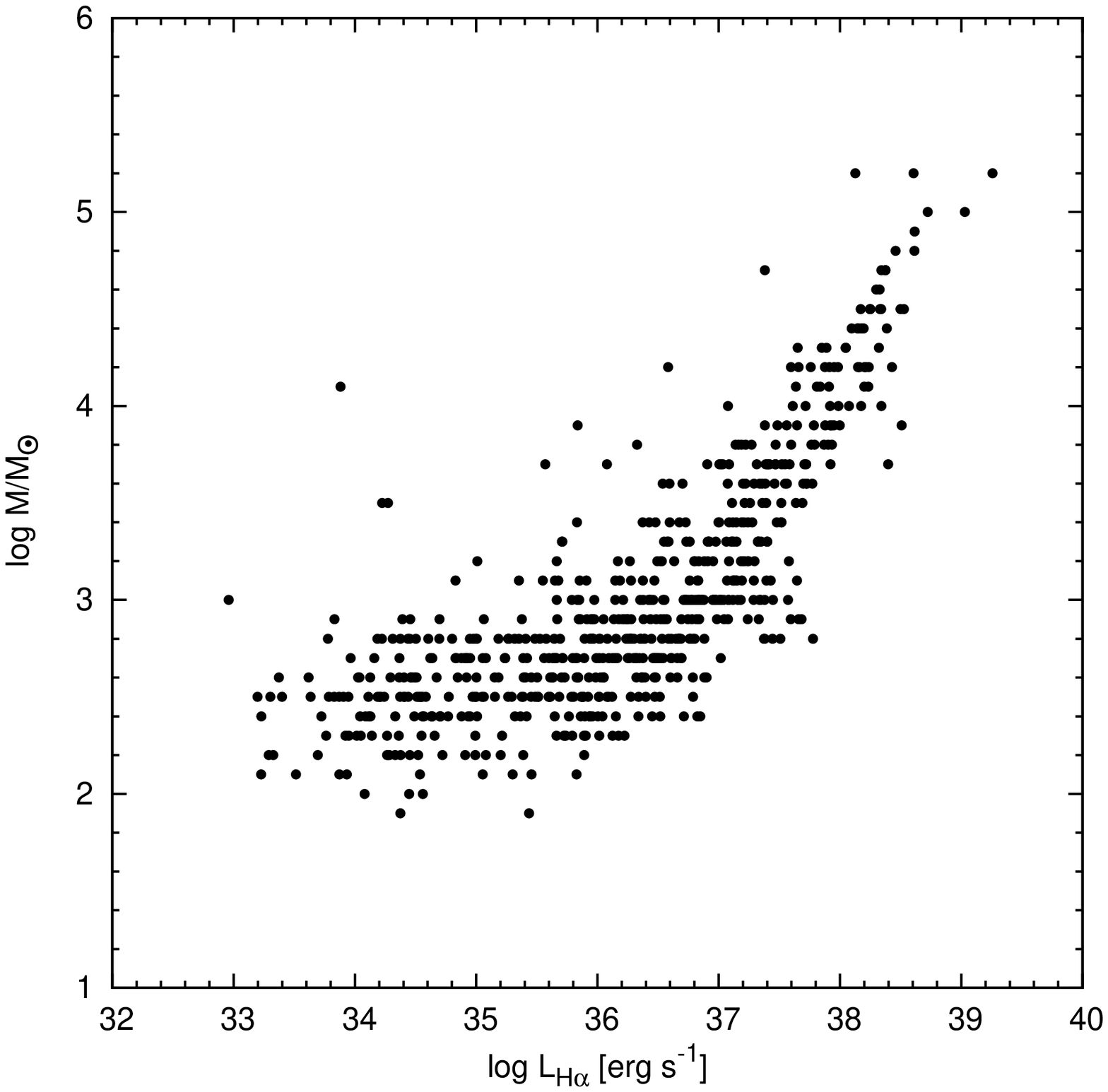}
\caption{\label{mr} {\it (Left panel)} Distribution of cluster mass as a function of 
size. The best fit for clusters with size $>10$\,pc is shown by a solid line.
Masses have been corrected for incompleteness.
{\it (Right panel)} Relation between cluster mass and the H$\alpha$ luminosity.  
}
\end{figure*}

The mass-radius relation of the YSCs of our sample is illustrated in the right panel of Fig. \ref{mr}.
Here, the size is the $24\,\mu$m semi-major axis from SExtractor.
The distribution shows a large scatter for sizes less than 10 pc,
as already noted for our Galaxy by \citet{1991ApJ...383..524Z} and  \citet{2004ASPC..322...19L} who
found a poor mass-radius correlation for compact star clusters (radius $<10$\,pc) with ages 
between a few tens to hundreds of Myr.
In nearby spiral galaxies, \citet{2004ASPC..322...19L} obtains $M \propto R^{2.1}$ 
in relatively young compact clusters with age $<1$\,Gyr and
$R<10$\,pc \citep[confirmed by][for YSC in M33]{2010A&A...521A..41G}.
In our sample YSCs larger than 10 pc show a good correlation with a log-log slope of $2.09 \pm0.01$ (least square fit).
In M33 \citet{2005ASSL..327..287B} obtain a similar slope considering giant molecular clouds ($M_{cl} \propto R^{2.2}$),
similar to other nearby galaxies such as M51 \citep{2005A&A...443...79B}  
for GMCs with similar ages than the YSCs in M33 but larger sizes. We may thus arrive at the
conclusion that the YSCs display average characteristics reminiscent of the
protocluster environment from which they were born.
A similar conclusion was also drawn by \citet{2010A&A...521A..41G} from the analysis of stellar clusters and complexes.

The relation between mass and H$\alpha$ luminosity is shown in the right panel of Fig. \ref{mr}. 
We observe a good correlation for cluster masses in excess of $\sim 500\,\msun$,
although the spread is significant at the faint end, as expected for the stochastic IMF
model (see previous Section).
 
The age distribution of the inner and outer YSCs is displayed in Fig. \ref{Hage}.
The  distribution of the former is narrower than the latter and indicate that
clusters tend to be younger in the inner regions of M33.
On the other hand, the cluster mass distribution for the two subsamples shown in Fig. \ref{Hmass}
reveals a broader spread for the inner sample, while the outer regions  peak at $\sim 10^2\,\msun$ and
are devoid of massive objects ($> 10^4\,\msun$). The distribution of
cluster masses, corrected for incompleteness of the IMF, is compatible with a 
differential distribution of index -2. This is consistent with the value found in the MW and in M33 \citep{1984AJ.....89.1822V,
1988AJ.....95..704C,2007ChJAA...7..155D}.

\begin{figure*}
\centering
\includegraphics[height=9cm,width=7cm,angle=-90]{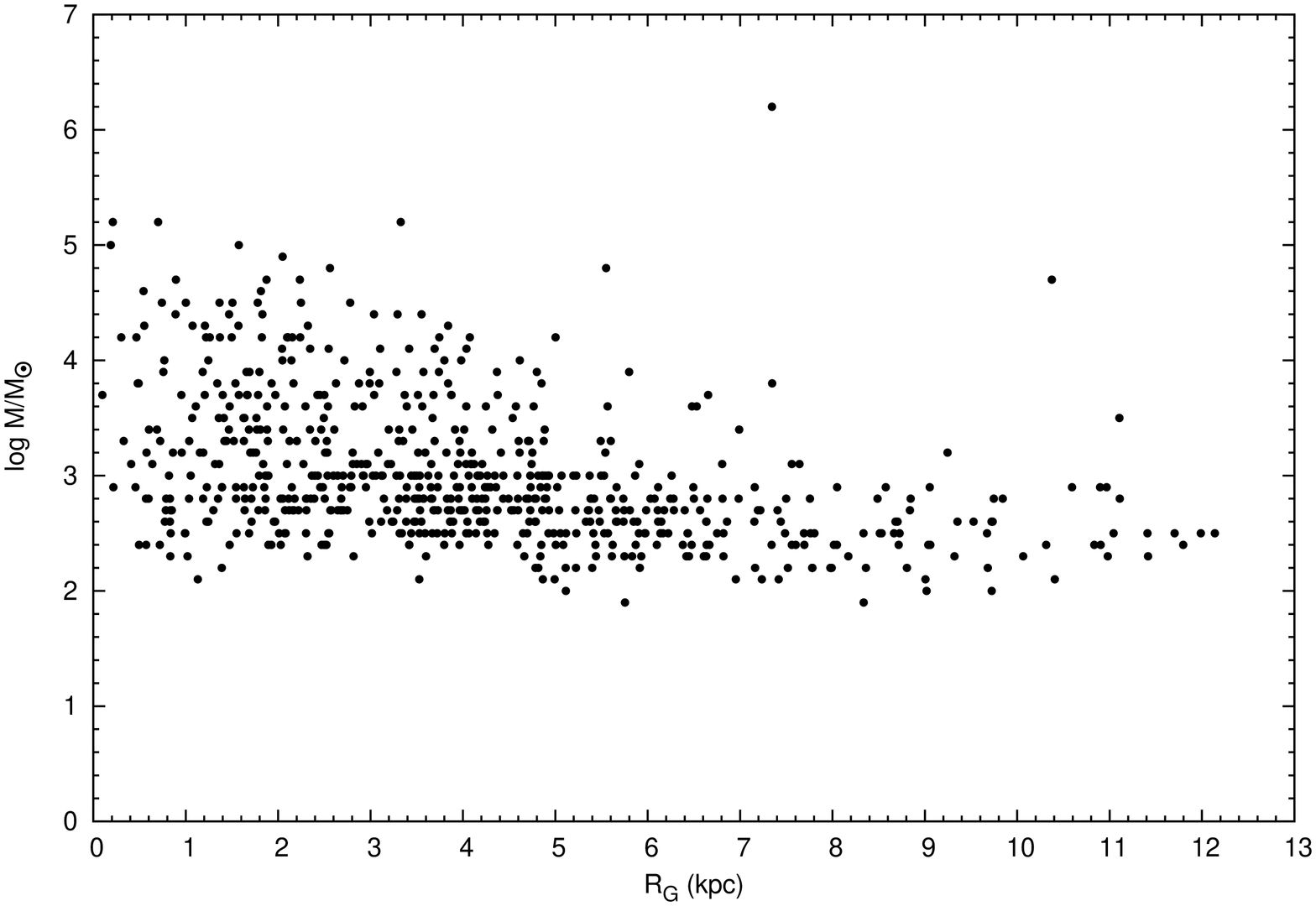}
\includegraphics[height=9cm,width=7cm,angle=-90]{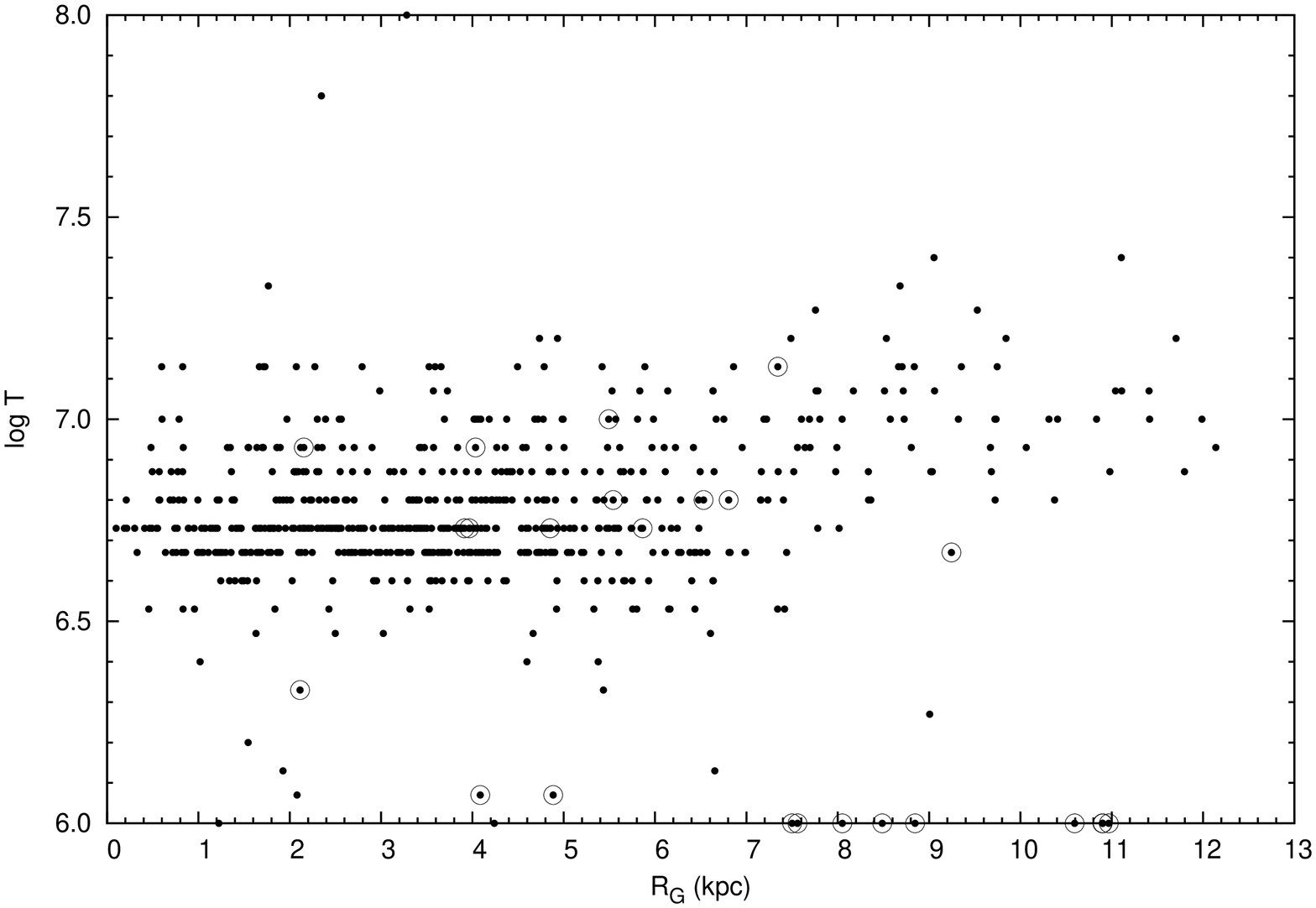}
\caption{\label{radial} The radial distribution of mass, age and $L_{TIR}/L_{FUV}$ of YSCs  as a function of the galactocentric distance $R_G$.
Circles indicate YSCs with unreliable fits ($\chi^2>100$).}
\end{figure*}

For the large majority of the YSCs selected in the mid-IR with a UV and H$\alpha$ counterpart
we find ages between 3 and 10\,Myr. Given the sample selection criteria, 
we are unable to trace the very early phases of the cluster formation process. 
Beyond the H$\alpha$ edge, at about 7\,kpc, 
we do not find clusters younger than 6\,Myr  and the average age is
of order 10\,Myr (see Figure~\ref{radial}). 
There is no clear evidence that this result is due to a selection bias as if, 
for example, outer YSCs,
being of smaller mass (and luminosity), would be more difficult to detect in
their early stages, especially due to the requirement of an H$\alpha$ and
UV counterpart. 
The lack of younger clusters in the outer regions is not ameliorated if the IMF is truncated at $40\,\msun$,
while this IMF model increases slightly the number of clusters with T$<3$~Myrs in the inner regions. 
 
Fig. \ref{Hav} shows the $A_V$ distributions. Most of the YSCs in our selected sample have A$_V$ between 0.2 and 1. 
We do not find evidence that the outer clusters suffer from lower extinction, although there is evidence for a
different extinction curve.
The radial variation of age and cluster mass for our sample of YSCs is shown 
in Fig. \ref{radial}.
We can clearly see that, with increasing galactocentric radii, the clusters become less  massive 
and  older. The SEDs of clusters at very large radii are characterized by
a higher IR-to-UV and IR-to-H$\alpha$ light ratios. This property is well fitted
by the synthesis models of older clusters. However, the presence of low-mass clusters
as young as $\sim$10\,Myr beyond the SF edge of the disk, where the average HI column disk is
only a few times 10$^{20}$~cm$^{-2}$, is remarkable. Localized density enhancements are evidently
present with the consequent formation of small H$_2$ clouds, birthplaces of low mass YSCs. The 
molecular gas dispersal time is comparable to the YSC ages that we find \citep{1990ApJ...359..319L}. Hence only
some residual gas  is present in the YSC environment,  similar to
that detected around MIR sources with H$\alpha$ counterparts close to the SF disk edge \citep{2011A&A...528A.116C}. 
Hence, the low extinction values found are due to our selection criteria: we have included in our sample
only MIR sources with FUV and H$\alpha$ counterparts.

\begin{figure*}
\centering
\includegraphics[height=6cm,width=6cm,angle=-90]{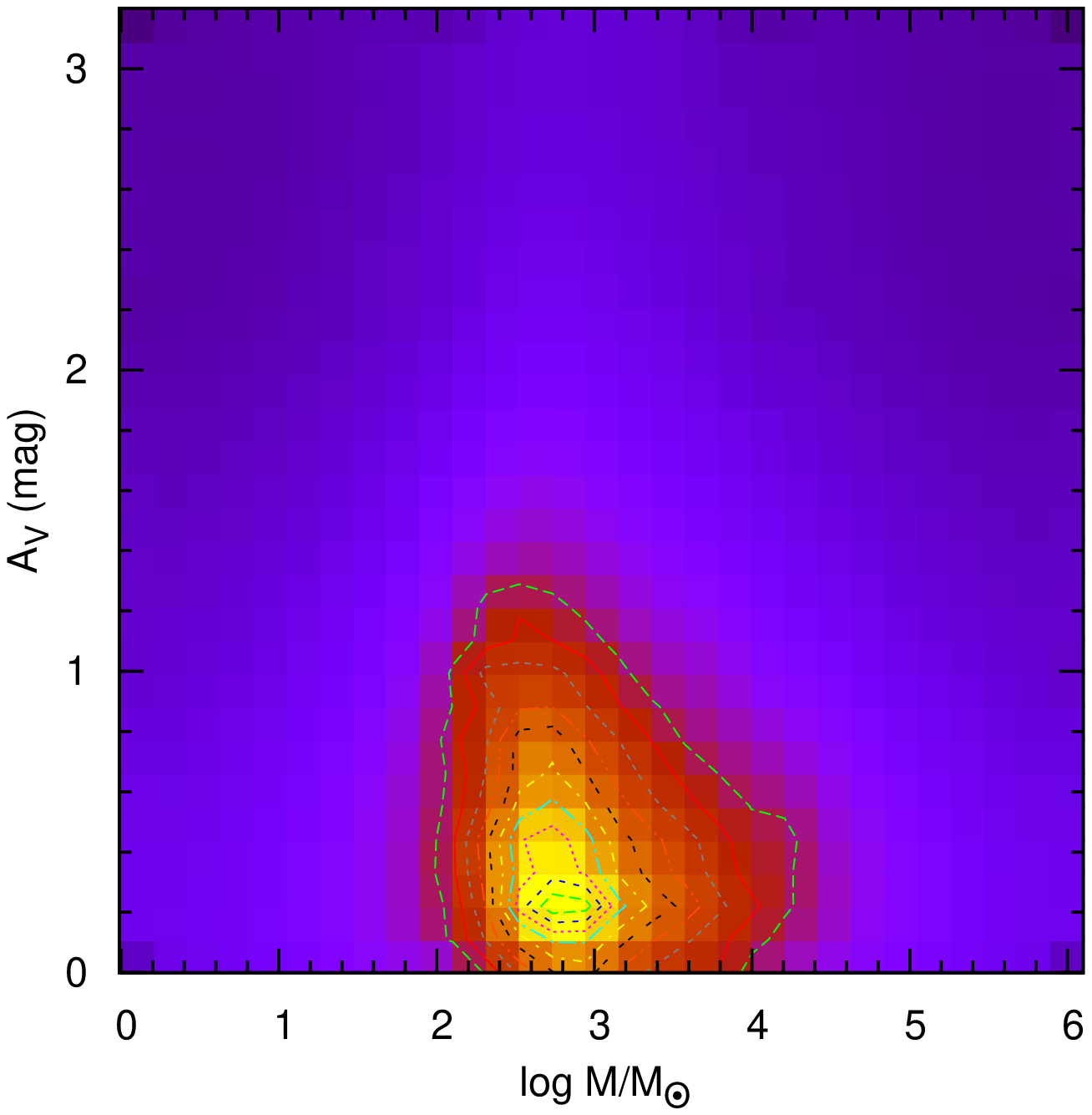}
\includegraphics[height=6cm,width=6cm,angle=-90]{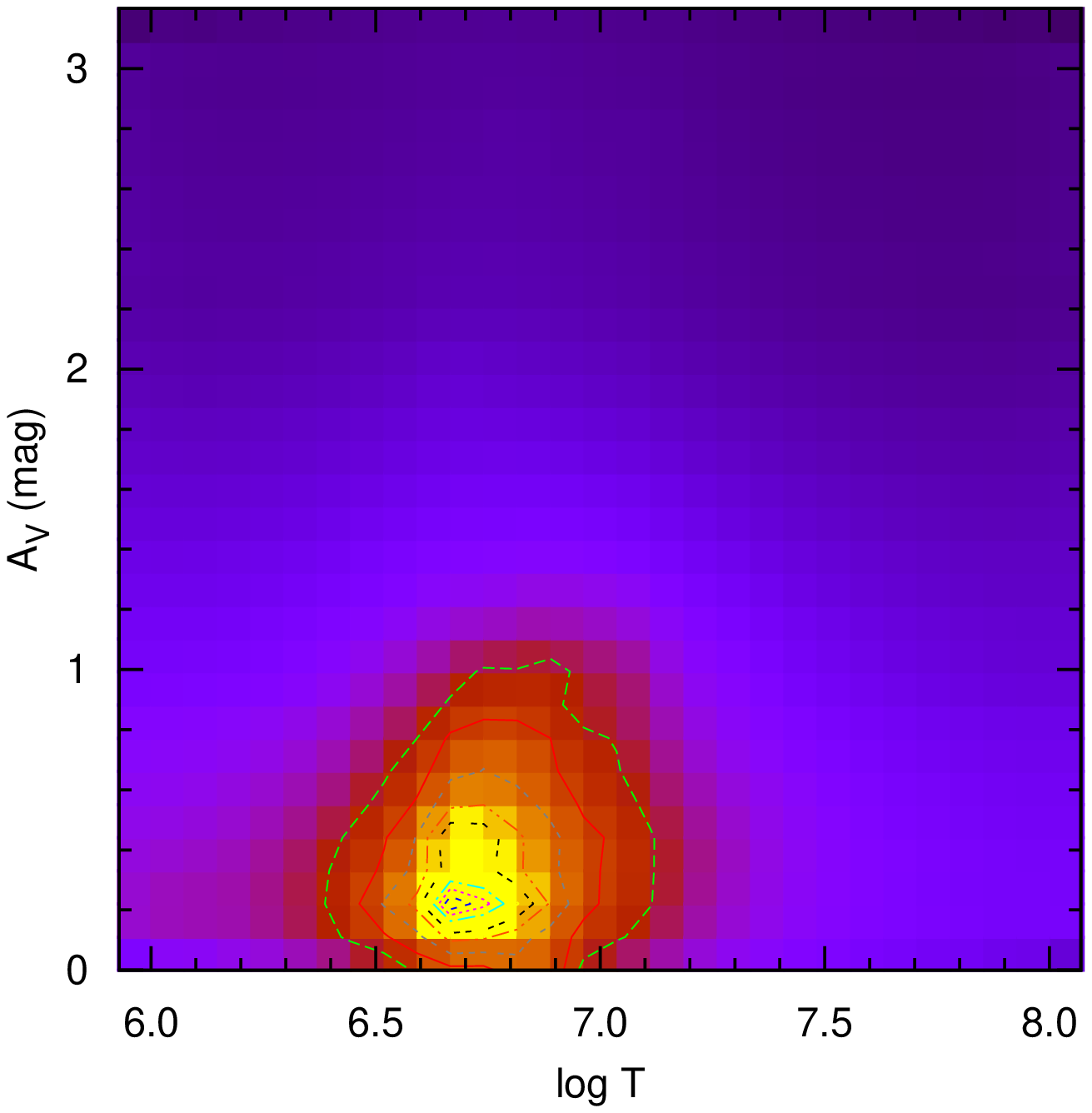}
\includegraphics[height=6cm,width=6cm,angle=-90]{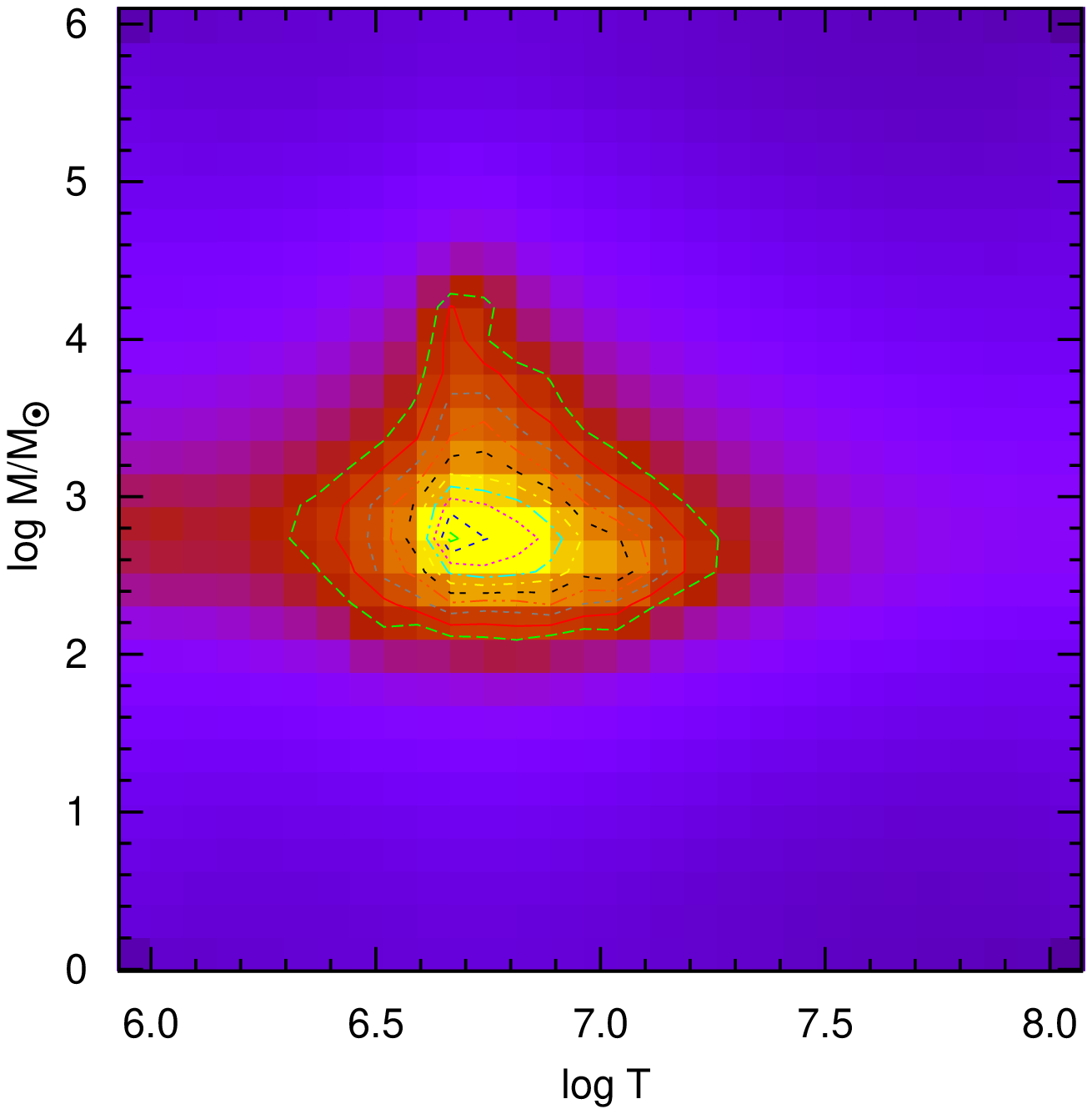}
\caption{\label{amr} Iso-density contours showing the age-mass-reddening relationship for the YSCs in our sample.
}
\end{figure*}

Finally, in Fig. \ref{amr} the relations between age, mass and $A_V$ are displayed as density images
to overcome the crowding of points. Again, we can see that the distributions are rather tight with most YSCs having
$A_V=$0-1\,mag, $log(T)=$6.5-7.1~Myr and $log(M/\msun)=$2-4.
The peak of the density maps corresponds to $A_V=0.3$\,mag, $log(T)=6.7$~Myr,
and $log(M/\msun)=2.5$.

\section{Summary and Conclusions}

In this study we have selected 915 mid-IR sources in M33
from the largest available $24\,\mu$m Spitzer map. 
Complementing the mid and far-infrared Spitzer data with
UV data from the GALEX satellite and with H$\alpha$ data,  
we have investigated the properties of 648 mid-IR sources which are YSC candidates 
distributed from small to large galactocentric radii. The observed SED for each YSC
is compared with theoretical synthesis models  to derive the cluster age, mass and extinction.
SED models with different leakage fractions, IMF upper mass cutoffs, extinction curves and 
dust absorption fraction in HII regions have been tested separately for the inner ($R_G<4$\,kpc) and 
outer cluster sample. This allows us to constrain the IMF and ISM properties of the inner and outer disk of M33.
The main results can be summarized as follows.

\begin{itemize}

\item
We find IR sources as far as 16\,kpc from the center, corresponding to the boundary of the 
$24\,\mu$m mosaic and to the extent of the warped HI disk. 
The completeness of the MIR source catalogue presented here is about 0.4~mJy, 
equivalent to the luminosity of a single B1.5V star. 
The radial decline of the IR source density becomes steeper at 
galactocentric distances $R_G=$4-5\,kpc and flattens out beyond the optical radius (8.5\,kpc).
YSCs selected in the mid-IR with H$\alpha$ and UV counterpart follow a similar radial distribution.

\item
The luminosity distribution of $24\,\mu$m sources as a function of the infrared 
or bolometric luminosity is steeper at the faint end in the outer regions, which are devoid of very bright sources.
The LF in the inner disk displays a double slope, 
steepening at the high luminosity end where
the slope is similar to that observed for HII regions in M33. 

\item
The average extinction of the YSCs is modest, since most of them have $A_V<1$.
The similarity between the bolometric and infrared LFs throughout the M33 disk and the lack of
radial trends in $A_V$ imply that the dusty environment does not vary appreciably between
star forming sites at small and large galactocentric radii. However, a Milky Way-type extinction curve
is preferred in the inner disk while for the outer disk an LMC2-type gives the best fitting SED models. 

\item
The mass-radius relation of YSCs shows a linear correlation with slope $2.09\pm0.01$,
in close agreement with the GMC mass-size relation. Hence, the stellar densities of the YSCs in our
sample are representative of the protocluster environment.

\item
Low luminosity clusters ($M<10^3\,\msun$) become the dominant 
population at large galactocentric radii.
We find YSCs with age of about 10\,Myr as far out as 12\,kpc, corresponding to the
extent of the H$\alpha$ map used for the identification of YSC candidates. 
Previous surveys have uncovered a population of $\sim$100\,Myr old stars 
in the outskirts of M33, but our finding implies a more recent episode of star formation 
in the outer galaxy.

\item 
The YSCs in the inner disk show a broad mass and age range, extending in mass up to
$10^5\,\msun$, and with ages between 3 and 10\,Myr.  
The peak of the YSC distribution lies at $A_V$ = 0.3 mag, age=5~Myr and mass= $300\,\msun$.

\item
The best fit synthesis models to the SED favour an IMF with an invariant 
upper mass cutoff at $100\,\msun$ throughout the M33 disk. 
Taking into account the stochasticity of the IMF, relevant for low-mass clusters which cannot
populate the IMF up to the upper mass end, 
we are able to reproduce the deviations from linearity of the   
UV-continuum--H$\alpha$ relation that we observe for YSCs. 

\end{itemize}

\begin{acknowledgements}
This work is based on observations made with the NASA Galaxy Evolution Explorer and with the $Spitzer$
Space Telescope. GALEX is operated for NASA by the California Institute of Technology under NASA contract 
NAS5-98034. The Spitzer Space Telescope is operated by the Jet Propulsion Laboratory, California Institute 
of Technology under a contract with NASA. We thank Rene' Walterbos for providing us the H$\alpha$ image of M33.
\end{acknowledgements}

\bibliography{mybib}{}

\begin{thebibliography}{60}
\expandafter\ifx\csname natexlab\endcsname\relax\def\natexlab#1{#1}\fi

\bibitem[{{Bastian} {et~al.}(2007){Bastian}, {Ercolano}, {Gieles},
  {Rosolowsky}, {Scheepmaker}, {Gutermuth}, \& {Efremov}}]{2007MNRAS.379.1302B}
{Bastian}, N., {Ercolano}, B., {Gieles}, M., {et~al.} 2007, \mnras, 379, 1302

\bibitem[{{Bastian} {et~al.}(2005){Bastian}, {Gieles}, {Efremov}, \&
  {Lamers}}]{2005A&A...443...79B}
{Bastian}, N., {Gieles}, M., {Efremov}, Y.~N., \& {Lamers}, H.~J.~G.~L.~M.
  2005, \aap, 443, 79

\bibitem[{{Bertin} \& {Arnouts}(1996)}]{1996A&AS..117..393B}
{Bertin}, E. \& {Arnouts}, S. 1996, \aaps, 117, 393

\bibitem[{{Blitz} \& {Rosolowsky}(2005)}]{2005ASSL..327..287B}
{Blitz}, L. \& {Rosolowsky}, E. 2005, in Astrophysics and Space Science
  Library, Vol. 327, The Initial Mass Function 50 Years Later, ed.
  {E.~Corbelli, F.~Palla, \& H.~Zinnecker}, 287--+

\bibitem[{{Caldwell} {et~al.}(1991){Caldwell}, {Kennicutt}, {Phillips}, \&
  {Schommer}}]{1991ApJ...370..526C}
{Caldwell}, N., {Kennicutt}, R., {Phillips}, A.~C., \& {Schommer}, R.~A. 1991,
  \apj, 370, 526

\bibitem[{{Calzetti}(2001)}]{2001PASP..113.1449C}
{Calzetti}, D. 2001, \pasp, 113, 1449

\bibitem[{{Calzetti} {et~al.}(2010){Calzetti}, {Chandar}, {Lee}, {Elmegreen},
  {Kennicutt}, \& {Whitmore}}]{2010ApJ...719L.158C}
{Calzetti}, D., {Chandar}, R., {Lee}, J.~C., {et~al.} 2010, \apjl, 719, L158

\bibitem[{{Cervi{\~n}o} \& {Luridiana}(2006)}]{2006A&A...451..475C}
{Cervi{\~n}o}, M. \& {Luridiana}, V. 2006, \aap, 451, 475

\bibitem[{{Christian} \& {Schommer}(1988)}]{1988AJ.....95..704C}
{Christian}, C.~A. \& {Schommer}, R.~A. 1988, \aj, 95, 704

\bibitem[{{Corbelli}(2003)}]{2003MNRAS.342..199C}
{Corbelli}, E. 2003, \mnras, 342, 199

\bibitem[{{Corbelli} {et~al.}(2011){Corbelli}, {Giovanardi}, {Palla}, \&
  {Verley}}]{2011A&A...528A.116C}
{Corbelli}, E., {Giovanardi}, C., {Palla}, F., \& {Verley}, S. 2011, \aap, 528,
  A116+

\bibitem[{{Corbelli} \& {Salucci}(2000)}]{2000MNRAS.311..441C}
{Corbelli}, E. \& {Salucci}, P. 2000, \mnras, 311, 441

\bibitem[{{Corbelli} \& {Schneider}(1997)}]{1997ApJ...479..244C}
{Corbelli}, E. \& {Schneider}, S.~E. 1997, \apj, 479, 244

\bibitem[{{Corbelli} {et~al.}(2009){Corbelli}, {Verley}, {Elmegreen}, \&
  {Giovanardi}}]{2009A&A...495..479C}
{Corbelli}, E., {Verley}, S., {Elmegreen}, B.~G., \& {Giovanardi}, C. 2009,
  \aap, 495, 479

\bibitem[{{Cox}(2000)}]{2000asqu.book.....C}
{Cox}, A.~N. 2000, {Allen's astrophysical quantities}, ed. {Cox, A.~N.}

\bibitem[{{Dale} \& {Helou}(2002)}]{2002ApJ...576..159D}
{Dale}, D.~A. \& {Helou}, G. 2002, \apj, 576, 159

\bibitem[{{Davidge} \& {Puzia}(2011)}]{2011arXiv1107.0077D}
{Davidge}, T.~J. \& {Puzia}, T.~H. 2011, ArXiv e-prints

\bibitem[{{Davidge} {et~al.}(2011){Davidge}, {Puzia}, \&
  {McConnachie}}]{2011ApJ...728L..23D}
{Davidge}, T.~J., {Puzia}, T.~H., \& {McConnachie}, A.~W. 2011, \apjl, 728,
  L23+

\bibitem[{{de Grijs} \& {Parmentier}(2007)}]{2007ChJAA...7..155D}
{de Grijs}, R. \& {Parmentier}, G. 2007, \cjaa, 7, 155

\bibitem[{{Elmegreen}(2006)}]{2006ApJ...648..572E}
{Elmegreen}, B.~G. 2006, \apj, 648, 572

\bibitem[{{Engargiola} {et~al.}(2003){Engargiola}, {Plambeck}, {Rosolowsky}, \&
  {Blitz}}]{2003ApJS..149..343E}
{Engargiola}, G., {Plambeck}, R.~L., {Rosolowsky}, E., \& {Blitz}, L. 2003,
  \apjs, 149, 343

\bibitem[{{Engelbracht} {et~al.}(2005){Engelbracht}, {Gordon}, {Rieke},
  {Werner}, {Dale}, \& {Latter}}]{2005ApJ...628L..29E}
{Engelbracht}, C.~W., {Gordon}, K.~D., {Rieke}, G.~H., {et~al.} 2005, \apjl,
  628, L29

\bibitem[{{Freedman} {et~al.}(1991){Freedman}, {Wilson}, \&
  {Madore}}]{1991ApJ...372..455F}
{Freedman}, W.~L., {Wilson}, C.~D., \& {Madore}, B.~F. 1991, \apj, 372, 455

\bibitem[{{Fumagalli} {et~al.}(2011){Fumagalli}, {da Silva}, \&
  {Krumholz}}]{2011arXiv1105.6101F}
{Fumagalli}, M., {da Silva}, R.~L., \& {Krumholz}, M.~R. 2011, ArXiv e-prints

\bibitem[{{Gil de Paz} {et~al.}(2007){Gil de Paz}, {Boissier}, {Madore},
  {Seibert}, {Joe}, {Boselli}, {Wyder}, {Thilker}, {Bianchi}, {Rey}, {Rich},
  {Barlow}, {Conrow}, {Forster}, {Friedman}, {Martin}, {Morrissey}, {Neff},
  {Schiminovich}, {Small}, {Donas}, {Heckman}, {Lee}, {Milliard}, {Szalay}, \&
  {Yi}}]{2007ApJS..173..185G}
{Gil de Paz}, A., {Boissier}, S., {Madore}, B.~F., {et~al.} 2007, \apjs, 173,
  185

\bibitem[{{Gil de Paz} {et~al.}(2008){Gil de Paz}, {Thilker}, {Bianchi},
  {Arag{\'o}n-Salamanca}, {Boissier}, {Madore}, {D{\'{\i}}az-L{\'o}pez},
  {Trujillo}, {Pohlen}, {Erwin}, {Zamorano}, {Gallego}, {Iglesias-P{\'a}ramo},
  {V{\'{\i}}lchez}, {Moll{\'a}}, {Mu{\~n}oz-Mateos}, {P{\'e}rez-Gonz{\'a}lez},
  {Pedraz}, {Sheth}, {Kennicutt}, \& {Swaters}}]{2008ASPC..396..197G}
{Gil de Paz}, A., {Thilker}, D.~A., {Bianchi}, L., {et~al.} 2008, in
  Astronomical Society of the Pacific Conference Series, Vol. 396, Formation
  and Evolution of Galaxy Disks, ed. {J.~G.~Funes \& E.~M.~Corsini}, 197--+

\bibitem[{{Gratier} {et~al.}(2010){Gratier}, {Braine}, {Rodriguez-Fernandez},
  {Schuster}, {Kramer}, {Xilouris}, {Tabatabaei}, {Henkel}, {Corbelli},
  {Israel}, {van der Werf}, {Calzetti}, {Garcia-Burillo}, {Sievers}, {Combes},
  {Wiklind}, {Brouillet}, {Herpin}, {Bontemps}, {Aalto}, {Koribalski}, {van der
  Tak}, {Wiedner}, {R{\"o}llig}, \& {Mookerjea}}]{2010A&A...522A...3G}
{Gratier}, P., {Braine}, J., {Rodriguez-Fernandez}, N.~J., {et~al.} 2010, \aap,
  522, A3+

\bibitem[{{Greenawalt} {et~al.}(1998){Greenawalt}, {Walterbos}, {Thilker}, \&
  {Hoopes}}]{1998ApJ...506..135G}
{Greenawalt}, B., {Walterbos}, R.~A.~M., {Thilker}, D., \& {Hoopes}, C.~G.
  1998, \apj, 506, 135

\bibitem[{{Grossi} {et~al.}(2010){Grossi}, {Corbelli}, {Giovanardi}, \&
  {Magrini}}]{2010A&A...521A..41G}
{Grossi}, M., {Corbelli}, E., {Giovanardi}, C., \& {Magrini}, L. 2010, \aap,
  521, A41+

\bibitem[{{Grossi} {et~al.}(2011){Grossi}, {Hwang}, {Corbelli}, {Giovanardi},
  {Okamoto}, \& {Arimoto}}]{2011arXiv1106.4704G}
{Grossi}, M., {Hwang}, N., {Corbelli}, E., {et~al.} 2011, ArXiv e-prints

\bibitem[{{Heyer} {et~al.}(2004){Heyer}, {Corbelli}, {Schneider}, \&
  {Young}}]{2004ApJ...602..723H}
{Heyer}, M.~H., {Corbelli}, E., {Schneider}, S.~E., \& {Young}, J.~S. 2004,
  \apj, 602, 723

\bibitem[{{Hoopes} \& {Walterbos}(2000)}]{2000ApJ...541..597H}
{Hoopes}, C.~G. \& {Walterbos}, R.~A.~M. 2000, \apj, 541, 597

\bibitem[{{Hunter} {et~al.}(2011){Hunter}, {Elmegreen}, {Oh}, {Anderson},
  {Nordgren}, {Massey}, {Wilsey}, \& {Riabokin}}]{2011arXiv1107.5587H}
{Hunter}, D.~A., {Elmegreen}, B.~G., {Oh}, S.-H., {et~al.} 2011, ArXiv e-prints

\bibitem[{{Kennicutt}(1989)}]{1989ApJ...344..685K}
{Kennicutt}, Jr., R.~C. 1989, \apj, 344, 685

\bibitem[{{Kroupa}(2001)}]{2001MNRAS.322..231K}
{Kroupa}, P. 2001, \mnras, 322, 231

\bibitem[{{Larsen}(2004)}]{2004ASPC..322...19L}
{Larsen}, S.~S. 2004, in Astronomical Society of the Pacific Conference Series,
  Vol. 322, The Formation and Evolution of Massive Young Star Clusters, ed.
  {H.~J.~G.~L.~M.~Lamers, L.~J.~Smith, \& A.~Nota}, 19--+

\bibitem[{{Larson}(1982)}]{1982MNRAS.200..159L}
{Larson}, R.~B. 1982, \mnras, 200, 159

\bibitem[{{Leisawitz}(1990)}]{1990ApJ...359..319L}
{Leisawitz}, D. 1990, \apj, 359, 319

\bibitem[{{Leitherer} {et~al.}(1999){Leitherer}, {Schaerer}, {Goldader},
  {Gonz{\'a}lez Delgado}, {Robert}, {Kune}, {de Mello}, {Devost}, \&
  {Heckman}}]{1999ApJS..123....3L}
{Leitherer}, C., {Schaerer}, D., {Goldader}, J.~D., {et~al.} 1999, \apjs, 123,
  3

\bibitem[{{Magrini} {et~al.}(2010){Magrini}, {Stanghellini}, {Corbelli},
  {Galli}, \& {Villaver}}]{2010A&A...512A..63M}
{Magrini}, L., {Stanghellini}, L., {Corbelli}, E., {Galli}, D., \& {Villaver},
  E. 2010, \aap, 512, A63+

\bibitem[{{Maraston} {et~al.}(2006){Maraston}, {Daddi}, {Renzini}, {Cimatti},
  {Dickinson}, {Papovich}, {Pasquali}, \& {Pirzkal}}]{2006ApJ...652...85M}
{Maraston}, C., {Daddi}, E., {Renzini}, A., {et~al.} 2006, \apj, 652, 85

\bibitem[{{McKee} \& {Williams}(1997)}]{1997ApJ...476..144M}
{McKee}, C.~F. \& {Williams}, J.~P. 1997, \apj, 476, 144

\bibitem[{{McQuinn} {et~al.}(2007){McQuinn}, {Woodward}, {Willner}, {Polomski},
  {Gehrz}, {Humphreys}, {van Loon}, {Ashby}, {Eicher}, \&
  {Fazio}}]{2007ApJ...664..850M}
{McQuinn}, K.~B.~W., {Woodward}, C.~E., {Willner}, S.~P., {et~al.} 2007, \apj,
  664, 850

\bibitem[{{Oey} \& {Clarke}(1998)}]{1998AJ....115.1543O}
{Oey}, M.~S. \& {Clarke}, C.~J. 1998, \aj, 115, 1543

\bibitem[{{Oke}(1990)}]{1990AJ.....99.1621O}
{Oke}, J.~B. 1990, \aj, 99, 1621

\bibitem[{{P{\'e}rez-Gonz{\'a}lez} {et~al.}(2006){P{\'e}rez-Gonz{\'a}lez},
  {Kennicutt}, {Gordon}, {Misselt}, {Gil de Paz}, {Engelbracht}, {Rieke},
  {Bendo}, {Bianchi}, {Boissier}, {Calzetti}, {Dale}, {Draine}, {Jarrett},
  {Hollenbach}, \& {Prescott}}]{2006ApJ...648..987P}
{P{\'e}rez-Gonz{\'a}lez}, P.~G., {Kennicutt}, Jr., R.~C., {Gordon}, K.~D.,
  {et~al.} 2006, \apj, 648, 987

\bibitem[{{Salpeter}(1955)}]{1955ApJ...121..161S}
{Salpeter}, E.~E. 1955, \apj, 121, 161

\bibitem[{{Schlegel} {et~al.}(1998){Schlegel}, {Finkbeiner}, \&
  {Davis}}]{1998ApJ...500..525S}
{Schlegel}, D.~J., {Finkbeiner}, D.~P., \& {Davis}, M. 1998, \apj, 500, 525

\bibitem[{{Thilker} {et~al.}(2007){Thilker}, {Bianchi}, {Meurer}, {Gil de Paz},
  {Boissier}, {Madore}, {Boselli}, {Ferguson}, {Mu{\~n}oz-Mateos}, {Madsen},
  {Hameed}, {Overzier}, {Forster}, {Friedman}, {Martin}, {Morrissey}, {Neff},
  {Schiminovich}, {Seibert}, {Small}, {Wyder}, {Donas}, {Heckman}, {Lee},
  {Milliard}, {Rich}, {Szalay}, {Welsh}, \& {Yi}}]{2007ApJS..173..538T}
{Thilker}, D.~A., {Bianchi}, L., {Meurer}, G., {et~al.} 2007, \apjs, 173, 538

\bibitem[{{Thilker} {et~al.}(2005){Thilker}, {Hoopes}, {Bianchi}, {Boissier},
  {Rich}, {Seibert}, {Friedman}, {Rey}, {Buat}, {Barlow}, {Byun}, {Donas},
  {Forster}, {Heckman}, {Jelinsky}, {Lee}, {Madore}, {Malina}, {Martin},
  {Milliard}, {Morrissey}, {Neff}, {Schiminovich}, {Siegmund}, {Small},
  {Szalay}, {Welsh}, \& {Wyder}}]{2005ApJ...619L..67T}
{Thilker}, D.~A., {Hoopes}, C.~G., {Bianchi}, L., {et~al.} 2005, \apjl, 619,
  L67

\bibitem[{{Vacca} {et~al.}(1996){Vacca}, {Garmany}, \&
  {Shull}}]{1996ApJ...460..914V}
{Vacca}, W.~D., {Garmany}, C.~D., \& {Shull}, J.~M. 1996, \apj, 460, 914

\bibitem[{{van den Bergh} \& {Lafontaine}(1984)}]{1984AJ.....89.1822V}
{van den Bergh}, S. \& {Lafontaine}, A. 1984, \aj, 89, 1822

\bibitem[{{V{\'a}zquez} \& {Leitherer}(2005)}]{2005ApJ...621..695V}
{V{\'a}zquez}, G.~A. \& {Leitherer}, C. 2005, \apj, 621, 695

\bibitem[{{Verley} {et~al.}(2009){Verley}, {Corbelli}, {Giovanardi}, \&
  {Hunt}}]{2009A&A...493..453V}
{Verley}, S., {Corbelli}, E., {Giovanardi}, C., \& {Hunt}, L.~K. 2009, \aap,
  493, 453

\bibitem[{{Verley} {et~al.}(2007){Verley}, {Hunt}, {Corbelli}, \&
  {Giovanardi}}]{2007A&A...476.1161V}
{Verley}, S., {Hunt}, L.~K., {Corbelli}, E., \& {Giovanardi}, C. 2007, \aap,
  476, 1161

\bibitem[{{Weidner} \& {Kroupa}(2004)}]{2004MNRAS.348..187W}
{Weidner}, C. \& {Kroupa}, P. 2004, \mnras, 348, 187

\bibitem[{{Weidner} \& {Kroupa}(2006)}]{2006MNRAS.365.1333W}
---. 2006, \mnras, 365, 1333

\bibitem[{{Weingartner} \& {Draine}(2001)}]{2001ApJ...548..296W}
{Weingartner}, J.~C. \& {Draine}, B.~T. 2001, \apj, 548, 296

\bibitem[{{Wyder} {et~al.}(1997){Wyder}, {Hodge}, \&
  {Skelton}}]{1997PASP..109..927W}
{Wyder}, T.~K., {Hodge}, P.~W., \& {Skelton}, B.~P. 1997, \pasp, 109, 927

\bibitem[{{Zepf} {et~al.}(1991){Zepf}, {Whitmore}, \&
  {Levison}}]{1991ApJ...383..524Z}
{Zepf}, S.~E., {Whitmore}, B.~C., \& {Levison}, H.~F. 1991, \apj, 383, 524

\end{thebibliography}
\bibliographystyle{aa}

\clearpage
\newpage

\begin{table*}[h]
\centering
\caption{\label{T515} The position, size and the galactocentric distances '$R_G$' for sources in  \citet{2007A&A...476.1161V}
classified as Young Stellar Clusters and included in the analysis of YSC properties presented in this paper.
The complete table is available in the electronic form only. }
\begin{tabular}{@{}lccrrcc@{}}
\hline
ID& $\alpha_{(2000)}$ & $\delta_{(2000)}$&Size & $R_G$   \\
  & (degree) & (degree) & (arcsec) & (kpc)  \\
\hline
     1& 23.374977& 30.309910&   15.3&  5.37 \\ 
     2& 23.361427& 30.290756&    3.8&  5.66 \\ 
     3& 23.372110& 30.301689&    7.1&  5.49 \\ 
     4& 23.332920& 30.305302&    5.1&  5.47 \\ 
     5& 23.335817& 30.304033&    3.6&  5.48 \\ 
     .&         .&         .&      .&   .   \\
     .&         .&         .&      .&   .   \\
     .&         .&         .&      .&   .   \\
     .&         .&         .&      .&   .   \\
   514& 23.599760& 30.950455&    7.5&  4.60 \\ 
 \hline
\end{tabular}
 
\end{table*}

\begin{table*}[h]
\centering
\caption{\label{T400} Additional 24~$\mu$m source catalogue (added to the catalogue of \citet{2007A&A...476.1161V}).
The position, size and the galactocentric distances '$R_G$' are given in addition to fluxes at 8 and $24\,\mu$m.
The  `-' symbol in column 7 and 8 indicate sources outside the $8\,\mu$m map boundary.
In the last column the symbol ``YSC'' indicate sources classified as Young Stellar Clusters and included in the analysis of YSC 
properties presented in this paper. The complete table is available in the electronic form only. }
\begin{tabular}{@{}lcccrcrcrcc@{}}
\hline
ID& $\alpha_{(2000)}$ & $\delta_{(2000)}$&Size & F$_{24}~\mu$m & $\sigma$&F$_8~\mu$m & $\sigma$ & $R_G$ & Note   \\
  & (degree) & (degree) & (arcsec) &(mJy)& (mJy)& (mJy)& (mJy)& (kpc)&   \\
 \hline
   516& 23.320433& 29.687447&   1.8&  1.072& 0.020&        -&     -&  15.31&       -  \\
   517& 23.397005& 29.691725&   1.6&  0.850& 0.016&        -&     -&  14.85&       -  \\
   518& 23.264629& 29.772163&   3.8& 15.433& 0.091&        -&     -&  14.53&       -  \\
   519& 23.341816& 29.774063&   1.2&  0.310& 0.014&        -&     -&  14.02&       -  \\
   520& 23.529587& 29.812027&   1.9&  1.118& 0.016&        -&     -&  12.60&       -  \\
   521& 23.348316& 29.812914&   1.8&  1.011& 0.017&        -&     -&  13.45&       -  \\
     .&         .&         .&     .&      .&     .&        .&     .&      .&       .  \\
     .&         .&         .&     .&      .&     .&        .&     .&      .&       .  \\
   783& 23.457613& 30.742470&   3.5&  1.779& 0.076&    3.918& 0.223&   1.35&     YSC  \\
   784& 23.308142& 30.744352&   3.5&  1.280& 0.057&    1.384& 0.109&   3.69&     YSC  \\
   785& 23.450867& 30.746641&   8.0& 17.984& 0.317&   36.865& 0.676&   1.47&     YSC  \\
   786& 23.823700& 30.746367&   1.1&  0.277& 0.007&    0.096& 0.014&   7.78&     YSC  \\
    .&         .&         .&     .&      .&     .&        .&     .&      .&        .  \\
    .&         .&         .&     .&      .&     .&        .&     .&      .&        .  \\
  915& 23.571716& 31.531706&   1.4&  0.336& 0.013&        -&     -&  13.74&        -  \\   
\hline
\end{tabular}
 
\end{table*}

\begin{table*}[h]
\centering
\caption{\label{Tlf} The slope of best least-square fit to the cumulative 
distribution of the clusters of inner and outer region of M33 galaxy.
The sample `a' ($24\,\mu$m sources (915)) and `d' (selected YSCs (648)) 
represents the  cumulative distribution of the clusters 
as a function of $24\,\mu$m flux whereas the sample `b' (8 $\mu$m sources (833)) and `c'  (selected YSCs (648))  represents the  cumulative 
distribution of the clusters as a function of $L_{TIR}$ and  $L_{bol}$ respectively.
The  $1^{st}$ and  $2^{nd}$ fit represents the fitting in the lower and higher flux (or luminosity) ranges.
}
\begin{tabular}{@{}cccc@{}}
\hline
 Sample & Region  & $1^{st}$ fit  &   $2^{nd}$ fit  \\
        &  & slope $\pm \sigma$ & slope $\pm \sigma$ \\
\hline
 (a)&inner &$-0.44\pm0.02$ &$-1.12\pm0.06$   \\
 $,,$&outer &$-0.84\pm0.03$ &$       -    $   \\
\\
 (b)&inner &$-0.42\pm0.02$ &$-1.38\pm0.11$   \\
 $,,$&outer &$-0.88\pm0.03$ &$       -    $   \\
\\
 (c)&inner &$-0.32\pm0.02$ &$-1.40\pm0.07$   \\
 $,,$&outer &$-0.68\pm0.04$ &$       -    $   \\
\\
 (d)&inner &$-0.38\pm0.02$ &$-1.12\pm0.06$   \\
 $,,$&outer &$-0.78\pm0.02$ &$       -    $   \\
\hline
\end{tabular}
\end{table*}

\begin{table*}[h]
\centering
\caption{\label{Tsum} The sum of the $\chi^2$ obtained for different set of model parameters for inner and outer disk clusters.}
\begin{tabular}{@{}clccrr@{}}
\hline
 IMF mass-  & $EXT_{curve}$ (MW/ & $K_{dust}$  & $F_{lkg}$ &  $\chi^2$ Inner     &  $\chi^2$ Outer \\
-cut off (M$_\odot)$      &LMC$_{AV}$/LMC2) & (\%) & (\%)) & (Mean) & (Mean) \\
\hline
100  &  MW         &  0 &  0 &   14.8   &     44.5  \\
100  &  LMC$_{AV}$ &  0 &  0 &   18.6   &     45.9  \\
100  &  LMC2       &  0 &  0 &   17.3   &     42.6  \\
100  &  MW         & 30 &  0 &   15.7   &     49.2  \\
100  &  LMC$_{AV}$ & 30 &  0 &   21.3   &     53.5  \\
100  &  LMC2       & 30 &  0 &   18.6   &     49.0  \\
100  &  MW         &  0 & 30 &   21.1   &     68.4  \\
100  &  LMC$_{AV}$ &  0 & 30 &   21.3   &     69.2  \\
100  &  LMC2       &  0 & 30 &   20.3   &     67.9  \\
100  &  MW         & 30 & 30 &   22.3   &     66.2  \\
100  &  LMC$_{AV}$ & 30 & 30 &   24.4   &     70.3  \\
100  &  LMC2       & 30 & 30 &   21.4   &     66.8  \\
 40  &  MW         &  0 &  0 &   15.4   &     50.7  \\
 40  &  LMC$_{AV}$ &  0 &  0 &   19.0   &     52.0  \\
 40  &  LMC2       &  0 &  0 &   17.3   &     48.0  \\
 40  &  MW         & 30 &  0 &   19.5   &     58.9  \\
 40  &  LMC$_{AV}$ & 30 &  0 &   25.8   &     65.2  \\
 40  &  LMC2       & 30 &  0 &   21.0   &     58.6  \\
 40  &  MW         &  0 & 30 &   29.1   &     82.4  \\
 40  &  LMC$_{AV}$ &  0 & 30 &   27.3   &     82.2  \\
 40  &  LMC2       &  0 & 30 &   26.4   &     80.9  \\
 40  &  MW         & 30 & 30 &   46.9   &     94.6  \\
 40  &  LMC$_{AV}$ & 30 & 30 &   52.5   &    102.1  \\
 40  &  LMC2       & 30 & 30 &   45.6   &     95.7  \\
\hline
\end{tabular}
\end{table*}

\clearpage
\label{lastpage}
\end{document}